\documentclass[aps,showpacs,prd,tightenlines,twocolumn,10pt,nofootinbib,amsmath,amssymb]{revtex4}

\usepackage[pdftex]{graphicx}
\usepackage[pdftex]{color}

\usepackage{verbatim}

\def\lsim{\mathrel{\raise.3ex\hbox{$<$\kern-.75em\lower1ex\hbox{$\sim$}}}}
\def\gsim{\mathrel{\raise.3ex\hbox{$>$\kern-.75em\lower1ex\hbox{$\sim$}}}}

\begin{document}

\title{A new stochastic approach to cumulative weak lensing}

\author{Kimmo Kainulainen} \email{kimmo.kainulainen@phys.jyu.fi}
\affiliation{Department of Physics, University of Jyv\"{a}skyl\"{a}, PL 35 (YFL), 
FIN-40014 Jyv\"{a}skyl\"{a}, Finland}
\affiliation{Helsinki Institute of Physics, University of Helsinki, PL 64, 
FIN-00014 Helsinki, Finland}

\author{Valerio Marra} \email{valerio.marra@jyu.fi}
\affiliation{Department of Physics, University of Jyv\"{a}skyl\"{a}, PL 35 (YFL), 
FIN-40014 Jyv\"{a}skyl\"{a}, Finland}
\affiliation{Helsinki Institute of Physics, University of Helsinki, PL 64, 
FIN-00014 Helsinki, Finland}

\begin{abstract}

We study the weak gravitational lensing effects caused by a stochastic distribution of dark matter halos. We develop a simple approach to calculate the magnification probability distribution function which allows us to easily compute the magnitude bias and dispersion for an arbitrary data sample and a given universe model.
As an application we consider the effects of single-mass large-scale cosmic inhomogeneities ($M\sim 10^{15} h^{-1} M_\odot$) to the SNe magnitude-redshift relation, and conclude that such structures could bias the PDF enough to affect the extraction of cosmological parameters from the limited size of present-day SNe data samples.
We also release {\tt turboGL} \cite{turbogl}, a simple and very fast ($\lesssim 1$s) Mathematica code based on the method here presented.

\end{abstract}


\maketitle

\section{Introduction}

Large-scale inhomogeneities are known to affect the light coming from very distant objects. It is important to understand these effects well if one is to use cosmological observations to accurately map the expansion history and determine the precise composition of the universe. In particular the evidence for dark energy in the current cosmolgical concordance model is heavily based on the analysis of the apparent magnitudes of distant type Ia supernovae (SNe)~\cite{Kim:2003mq,Kowalski:2008ez}.
One way inhomogeneities can affect the observed SNe magnitude-redshift relation is through gravitational lensing.
How large these effects are depends strongly on the assumptions regarding size and density contrast of the structures through which light passes on its way from source to observer. The main purpose of this paper is to develop a simple tool to compute weak lensing effects on the SNe magnitude-redshift relation due to statistical distributions of large-scale inhomogeneities. 

The effect of inhomogeneities on cosmolgical observables have been studied earlier by many authors and in many different contexts. The effect of matter clumping into isolated halos, or isolated cores, was already considered by Kantowski in 1969~\cite{Kantowski:1969}, and more recent analyses of gravitational lensing by statistically distributed inhomogeneities have been carried out, for example, in Refs.~\cite{Wambsganss:1996qc, Holz:1997ic,Tomita:1999tf, Tomita:1999tg, Bergstrom:1999xh,Jain:1999ir,Vale:2003ad,Holz:2004xx,Martel:2007fh,Yoo:2007bt,Vanderveld:2008vi, Brouzakis:2007zi}.
Lately many authors have studied exactly solvable models for large-scale inhomogeneities, such as swiss-cheese, onion and meatball models~\cite{Kantowski:1969,Marra:2007pm,Marra:2007gc,Biswas:2007gi,Biswas:2006ub,Marra:2008sy,Bolejko:2008xh,Valkenburg:2009iw,Kainulainen:2009sx,Clifton:2009jw}.
The local Hubble bubble scenario has also been quite succesfull in explaining many of the cosmological observations~\cite{Mustapha:1998jb, celerier, Tomita:2000jj, Iguchi:2001sq, alnes0602, Chung:2006xh, Enqvist:2006cg, Alexander:2007xx, GarciaBellido:2008yq,Zibin:2008vk}, obviously with the price of giving up the cosmological principle. Yet another aspect of inhomogeneities is to cause possibly strong backreaction effects on the dynamical Einstein equations governing the evolution of the metric~\cite{Kolb:2009rp,ellis,Buchert:2007ik,Kolb:2005da,Notari:2005xk,Coley:2005ei,Rasanen:2006kp,Kai:2006ws,Paranjape:2007wr,Wiltshire:2007jk,Kolb:2008bn,Bolejko:2008yj,Brown:2009tg,Larena:2008be,Clarkson:2009hr}.

The approach in this paper is closest in spirit to that of Refs.~\cite{Holz:1997ic, Holz:2004xx}. Here the cosmological principle is respected and we will assume that the inhomogeneities can be introduced as a perturbation on a well defined global background solution. We will also neglect all redshift effecs in voids (see Refs.~\cite{Biswas:2006ub, Marra:2007pm, Marra:2007gc, Brouzakis:2008uw}) and concentrate purely on cumulative weak lensing~\cite{Bartelmann:1999yn}. The central physical concept for the present work is the observation, made already by Zel'dovich in 1964~\cite{zel}, that light travelling in empty parts of a clumpy but globally homogenous universe becomes demagnified.
Light going through mass concentrations is magnified on the other hand and the null result of the Friedmann-Lema\^{i}tre-Robertson-Walker (FLRW) background solution only arises through averaging over a large number of photon geodesics. As a result, the magnification probability distribution function (PDF) for a single observation is skewed favouring mild demagnifications, with a long compensating tail of positive magnifications. Such skewness is a concern for the interpretation of the cosmological data, because the demagnification effect could be misinterpretated as a relative dimming of standard candles in small data sets. The degree of skewness of the PDF obviously depends on the size and the spatial distribution of the matter concentrations.

The central object to compute then is the magnification probability distribution function for a random photon geodesic in a given inhomogenous universe model. Refs.~\cite{Holz:1997ic, Holz:2004xx} evaluated the PDF numerically performing large simulations in a model universe and computing the magnification factors with ray-tracing techniques. Here we will introduce a much simpler way to compute the PDF and give an explicit expression for it as a sum over probability weighted configurations of inhomogenities. Moreover, we will develop a simple analytic approximation that can reproduce the mode and the dispersion of the numerical PDF. In addition to the statistical bias due to skewness, there may be systematic biases on the observed PDF, such as extinction, foreground light contamination, strong lensing and outlier corrections.
These biases can be problematic because they, unlike the statistical magnification bias, persist even in large data sets. We will not try to estimate the size the systematic biases here, but we will show how their effects can easily be included in the analysis.

We test our method by reproducing the key results of Refs.~\cite{Holz:1997ic, Holz:2004xx} for a universe in which all matter is homogenously distributed in halos with $M \sim 10^{12} h^{-1} M_\odot$ and our results are found to be in good agreement. We also consider much larger halos of size $M \sim 10^{15} h^{-1} M_\odot$, whose existence is suggested by the large voids and filamentary structures seen both in the large-scale simulations and in the galaxy redshift surveys~\cite{BoylanKolchin:2009nc,Fosalba:2007mf,surveys}.  From the weak lensing point of view such halos can be considered as localized lenses, irrespectively of whether they are gravitationally bound or not.
We again produce the simulated PDFs for a universe with such large structures, both in the $\Lambda$CDM and in the Einstein-de Sitter (EdS) background model.
We also produce the distributions for binned sets of observations and compute bias and dispersion for these effective PDFs.
Quite interestingly, supercluster size halos with $M \sim 10^{14} h^{-1} M_\odot$ turn out to represent roughly the borderline for when the skewness effects become important in the PDFs for the current best data sets.

This paper is organized as follows. In Section  \ref{setup} we define the background spacetime and the precise form and distribution of the inhomogeneities, and review the basic formalism needed to compute the weak lensing convergence. In Section \ref{wlcal} we develop the probabilistic formalism to calculate the lensing magnification PDF. In Section \ref{sec:analytic} we derive the analytic approximation for the lensing PDF and its mode and dispersion. In Section \ref{sec:results} we apply our methods to compute the PDFs for both the $\Lambda$CDM model and the EdS model for different types of matter distributions, and compare our numerical and analytical results. In Section \ref{conclusions} we will give our conclusions. In Appendix \ref{weds} we give some analytical results for the EdS model and in Appendix \ref{identity} we prove an identity used in the calculations.

%
\section{Setup} 
\label{setup}
%

In order to concentrate purely on the effects of weak gravitational lensing, we will ignore a possible strong backreaction (see Ref.~\cite{Kolb:2009rp} for a definition and, for example, Ref.~\cite{ellis,Buchert:2007ik,Kolb:2005da,Notari:2005xk,Coley:2005ei,Rasanen:2006kp,Kai:2006ws,Paranjape:2007wr,Wiltshire:2007jk,Kolb:2008bn,Bolejko:2008yj,Brown:2009tg,Larena:2008be,Clarkson:2009hr} for a discussion). In particular we will assume that the spacetime of the inhomogeneous universe is accurately described by small perturbations around the FLRW solution whose energy content and spatial curvature are defined as Hubble-volume spatial averages over the inhomogeneous universe. Following Ref.~\cite{Kolb:2009rp} we will call this Hubble-volume average the Global Background Solution (GBS), while the cosmological background solution actually obtained through the observations will be called the Phenomenological Background Solution (PBS).
The inhomogeneities we will introduce will cause the PBS depart from the GBS, and so this work falls into the category of weak backreaction. Also other phenomena, such as a local Hubble bubble~\cite{Mustapha:1998jb, celerier, Tomita:2000jj, Iguchi:2001sq, alnes0602, Chung:2006xh, Enqvist:2006cg, Alexander:2007xx, GarciaBellido:2008yq,Zibin:2008vk} or redshift effects~\cite{Biswas:2006ub, Marra:2007pm, Marra:2007gc, Brouzakis:2008uw} could be studied within the weak backreaction scenario. In a companion paper~\cite{Kainulainen:2009sx} we have extended the current work to include a local Hubble bubble, but here we will ignore this possibility and we also neglect all redshift effects~\cite{Kantowski:1969}.
In what follows, we will first introduce the GBS and the detailed form of the inhomogeneities. Then, after defining the parameters that describe our model universe, we will briefly review the machinery necessary to calculate weak-lensing effects.

%
\subsection{Global background solution} 
\label{bkg}
%

In agreement with CMB observations we will focus on spatially flat models. Moreover, since we are only interested on the late evolution of the universe ($z \leq 1.6$) we neglect radiation retaining only the contributions from (dark and baryonic) matter and the dark energy in the form of a cosmological constant. The parameters that specify the GBS will be therefore be $\Omega_{M, 0}$ and $H_{0}$. As special cases we will consider in particular the $\Lambda$CDM model with $\Omega_{M, 0}=0.28$ and the EdS model with $\Omega_{M, 0}=1$. The evolutions of the Hubble expansion rate and of the density parameters as a function of redshift are given by
\begin{eqnarray}
H(z)&=& H_{0} \left( \Omega_{M, 0} \, (1+z)^{3}+\Omega_{\Lambda, 0} \right)^{1/2} \label{hhh} \\
\Omega_{M}(z)&=& \Omega_{M,0} \, (1+z)^{3} \frac{H^2_0}{H^2(z)} \label{om1} \\
\Omega_{\Lambda}(z)&=&1-\Omega_{M}(z)\,,
\end{eqnarray}
where the subscript $0$ will denote the present-day values of the quantities throughout this paper and $H_{0}=100  h$ km s$^{-1}$ Mpc$^{-1}$. Substituting in Eq.~(\ref{hhh}) $H=\dot{a}(t)/a(t)$ and $1+z=a_{0}/a(t)$ we obtain the equation we have to solve in order to find the time evolution of the GBS.
The observables we will be interested in are the angular diameter distance, the luminosity distance and the distance modulus:
\begin{eqnarray}
D_A(z)&=&{c \over 1+z}\int_{0}^{z}{d\bar{z} \over H(\bar{z})} \\
D_L(z)&=& (1+z)^{2}D_A(z) \\
m(z)&=& 5 \log_{10}{D_L(z) \over 1\textrm{Mpc}} + 25 \,.
\end{eqnarray}
Analytical expressions for these quantities in the EdS case can be found in Appendix \ref{weds}. 

\subsection{Inhomogeneities}
\label{sec:inohomogenities}

We will first describe the statistical distribution of the inhomogeneities within the GBS discussed in the previous section. The more precise properties of the inhomogeneities, or halos, will be described in the second subsection and  in the last subsection we summarize the parameters that describe our model universe.

\subsubsection{General statistical properties} \label{sec:inohomogenities1}

The inhomogeneity scale introduced by the halos will be much smaller than the Hubble radius, $d_{c} \ll c/H_{0}$, so that the GBS can perturbatively describe the spacetime.
In particular we will assume that redshifts can be related to co-moving distances through the GBS. In our setup the halos are randomly distributed with a co-moving number density $n_c$ and all the matter in the universe is within these objects (see Eq.~\ref{m1}).
We will later generalize this picture for a continuous mass distribution of halos.

\begin{figure}
\begin{center}
\includegraphics[width=8.5cm]{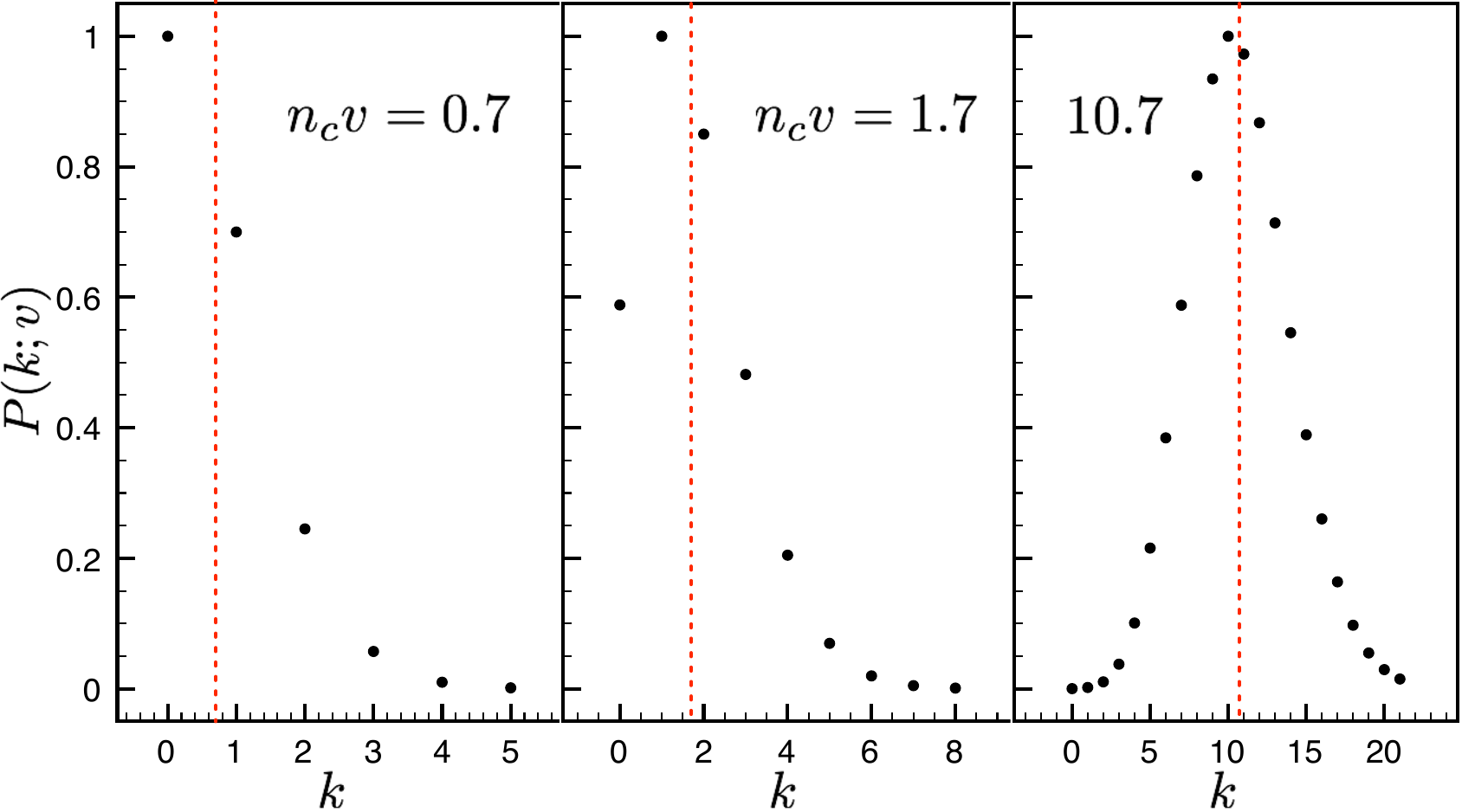}
\caption{Poisson distribution with maximum normalized to unity for four different values of $n_{c}v$. The vertical dotted lines mark the mean.}
\label{Pk}
\end{center}
\end{figure}

Let us now consider a co-moving volume $V$ much larger than the characteristic scale $n_c^{-1}$ so that the total number $N_{H}$ of halos in $V$ is large, $N_{H} \simeq n_c V \gg 1$.  Let us first estimate the average co-moving distance $d_{c}$ between the halos. To this end we need the probability $P(k;v)$ of having $k$ objects in the co-moving volume $v$:
\begin{eqnarray} 
P(k; v)&=&\binom{N_{H}}{k} \left( {v \over V} \right)^{k} \left( 1-{v \over V} \right)^{N_{H}-k}
\nonumber \\
&=&\binom{N_{H}}{k} \left( {n_{c}v \over N_{H}} \right)^{k} \left( 1-{n_{c}v \over N_{H}} \right)^{N_{H}-k}
\nonumber \\
&\stackrel{N_{H} \rightarrow \infty}{\longrightarrow}&
e^{- n_{c}v}{(n_{c}v)^{k} \over k!} \,,
\end{eqnarray}
where we first wrote the binomial probability for $p=v/V$ and then took the limit $N_{H} \gg 1$.
We have thus found that the number of halos in the volume $v$ is distributed as a Poisson random variable of parameter $n_{c}v$ which is, as expected, the mean number of halos.
In particular, mean, mode, variance and skewness functions characterizing the distribution are:
\begin{eqnarray}
\bar{k}&=& n_{c}v  \nonumber \\
\hat{k}&=& \lfloor n_{c}v \rfloor \nonumber \\
\sigma&=& (n_{c}v)^{1/2} \nonumber \\
\gamma_{1} &=& (n_{c}v)^{-1/2} \,,
\label{kkk}
\end{eqnarray}
where the notation $\lfloor x \rfloor$ refers to the floor function corresponding to the largest integer not greater than $x$ and $\gamma_{1}$ is the skewness. These are of course well known statistical properties of the Poisson distribution. However, the skewness of the distribution is so crucial to our results that we illustrate this behaviour as a function of $n_{c}v$ in Fig.~\ref{Pk}. It is evident that the distribution is strongly skewed for small $n_{c}v$, whereas for increasing $n_cv$ the mode starts to approach the mean (marked by a vertical dotted line), skewness goes to zero and the distribution approaches the normal distribution.
In particular the difference between mode and mean for low values of $n_{c}v$ will be of central relevance for us in what follows.

Let us now return to the evaluation of $d_c$. Evidently $n_{c}V$ is a stable expectation for the number of objects in the volume $V$. To see this imagine that the volume $V$ is embedded within an even bigger volume $W$. The probability of having $k$ objects within $V$ is then given by the Poisson probability $P(k; V)$, but because $n_{c}V \gg 1$ this is is well approximated by a normal distribution with mean $\bar{k}=n_{c}V$ and variance $\sigma=(n_{c}V)^{1/2}$. For $n_{c}V \rightarrow \infty$ we therefore obtain $\sigma/\bar{k}\rightarrow 0$; that is, the relative fluctuations around the expected value will go to zero and we can simply write $N_{H} = n_{c}V$.
The latter property that the Poisson distribution tends to a delta function for large parameters will be important when we will discuss the lensing bias with respect to the size of the data samples of observations.

Because the objects are randomly distributed, the probability $w(r)dr$ that the nearest neighbour to an object is at a radial distance between $r$ and $r+dr$ equals the probability of having no particle within $r$ times the probability of having at least one particle between $r$ and $r+dr$:
\begin{eqnarray}
w(r)dr &=& P(0; v)P(\geq1; dv) 
\nonumber \\
&=& P(0; v)(1-P(0; dv))
\nonumber \\
&=& e^{- n_{c}v}(1-e^{- n_{c}dv}) = n_{c} \, dv \, e^{- n_{c}v} \,,
\end{eqnarray}
where $v={4 \pi r^{3} / 3}$. Therefore, the average co-moving distance $d_{c}$ between nearest neighbours is~\cite{Chandrasekhar:1943ws}:
\begin{eqnarray}
d_{c} &=& \int_{0}^{r_{max}}r \, w(r)dr 
={\Gamma(4/3) \over (4 \pi /3)^{1/3}} \, n_{c}^{-1/3} 
\nonumber \\
&\simeq& 0.55 \, n_{c}^{-1/3} \,,
\label{eq:nc}
\end{eqnarray}
where we put $r_{max}=\infty$ using the fact that $V\gg n_{c}^{-1}$.

\subsubsection{Detailed description of the halos} \label{deschalo}

We take our halos to be spherical and characterized by a co-moving radius $R$ where the density of the halo goes to zero, so that beyond this distance a halo does not contribute to the weak lensing convergence. The density profile within the radius $R$ can be taken to be any smooth function like a gaussian (in which case $R\simeq 3 \sigma$), a singular isothermal sphere (SIS) or the Navarro-Frenk-White (NFW) profile \cite{Navarro:1995iw}.
All matter is taken to be in these halos, and so the distribution of matter within the co-moving volume $V$ is:
\begin{equation}
\rho_{M}= \sum_{j=1}^{N_{H}} \rho_{M j} = \bar{\rho}_{M} \sum_{j=1}^{N_{H}}\varphi(|r-r_{j}|) \,,
\label{inhomog1}
\end{equation}
where $\bar{\rho}_M=\bar{\rho}_{M, 0}(1+z)^{3}$ with $\bar{\rho}_{M, 0} =  
3 H_0^2 \Omega_{M,0}/(8 \pi G)$. Throughout this paper we use the overbar to denote a quantity corresponding to the GBS. The halo profile $\varphi(r)$ must be normalized appropriately to get the average density. From Eq.~(\ref{inhomog1}) we find:
\begin{eqnarray} 
\label{norma}
   1   &=&   \frac{1}{V}\int_{V} \sum_{j=1}^{N_{H}} \varphi_{j} \, dV 
    \simeq \frac{N_{H}}{V} \int_{V} \varphi \, dV
\nonumber \\
&\Longrightarrow& 
  \int_{V} \varphi \, dV  =  {1 \over n_c} \,,
\label{eq:normof-f}
\end{eqnarray}
where $\varphi_j\equiv \varphi(|r-r_j|)$ and in the second equality we used the fact that $N_{H}\gg 1$ in order to neglect boundary corrections. Equation (\ref{norma}) ties the halo mass to the co-moving halo number density:
\begin{equation}
M_{H} =  \bar{\rho}_{M} a^{3} \int_{V}\varphi \, dV =  {\bar{\rho}_{M,0} \, a_{0}^{3}\over  n_{c}} 
\label{m1} 
\end{equation}
and $n_c$ is further related to the average distance $d_{c}$ between halos through Eq.~(\ref{eq:nc}).
In the present analysis $n_{c}$ is a constant with redshift:
we will extend this picture in Section~\ref{sec:fm}.
The total energy content of the universe and its density contrast can now be expressed as:
\begin{eqnarray}
\rho&=&\bar{\rho}_{M} \sum_{j=1}^{N_{H}}\varphi_{j}+\rho_{\Lambda} 
\\
 \delta&=&{\delta\rho\over \bar{\rho}} 
= {\rho_{M} - \bar{\rho}_{M} \over \bar{\rho}} = \Omega_{M} \, \delta_{M}
\nonumber \\
&=&  \Omega_{M} \Big( \sum_{j=1}^{N_{H}}\varphi_{j}-1 \Big)
= \delta_{H}+\delta_{E} \,,
\label{deltaeq}
\end{eqnarray}
where $\delta_{M}=\rho_{M} / \bar{\rho}_{M}-1$ and
we have defined $\delta_{H}\equiv\Omega_{M}\sum_{j=1}^{N_{H}}\varphi_{j}$ and $\delta_{E}\equiv\Omega_{\Lambda}-1=-\Omega_{M}$.
The latter gives the density contrast in the empty space (voids) between the halos, while the former the matter field due to the halos.
The average contrast of a halo is found by averaging $\delta_{H}$ over the volume of a halo:
\begin{equation}
\langle \delta_{H} \rangle (z)= {\Omega_{M}(z) \over  n_{c} {4 \pi \over 3} R^{3}(z)} -1 \, .
\label{m2} 
\end{equation}
Eq.~(\ref{m2}) will be used to relate the truncation radius $R$ to the average contrast at virialization.

We stress at this point that by ``halo'' we do not mean only gravitationally bound systems, but also non virialized large-scale structures for which the radius $R$ is not related to the virialization contrast of Eq.~(\ref{m2}) (see Section \ref{sec:cdh}).

\subsubsection{Parameters of the model universe}

Summarizing, the parameters that specify our model universe are the matter abundance $\Omega_{M,0}$ and the Hubble expansion rate $H_{0}$ giving the GBS and the scales $R$ and $n_{c}$ describing the inhomogeneities.
Another important parameter is the number of supernova observations $N_O$ at a given redshift.
As we will see, because of lensing effects, the PBS will depend on $N_{O}$ and it will reduce to the GBS only when the observations cover uniformly the entire sky in the limit $N_O \rightarrow \infty$. The latter limit can also be understood as averaging over all the sky.
In Section  \ref{sec:observations} we will discuss this topic in detail including a redshift dependent $N_{O}(z)$-function. In Section ~\ref{sec:selection} we will consider the case where the observations fail to cover the entire sky and discuss how these selection effects may affect the PBS.
Thus, in what follows we will study quantitatively the lensing predictions for universes described by the parameter sets $(\Omega_{M,0}, H_0, R, n_c, N_O)$.

\subsection{Cumulative gravitational weak lensing}
\label{sec:cumulative}

We will now briefly introduce the tools necessary to calculate weak-lensing effects for our setup. For more details see Ref.~\cite{Bartelmann:1999yn}. In the weak-lensing theory the net magnification $\mu$ produced by a localised density perturbation is:
\begin{equation} \label{mag}
\mu={1 \over (1-\kappa)^{2} - |\gamma_{s}|^{2}}
\simeq (1-\kappa)^{-2}\,,
\end{equation}
where $\kappa$ is the lens convergence and $\gamma_{s}$ is the shear which we assume to be negligible~\cite{Bartelmann:1999yn}. The shift in the distance modulus caused by $\mu$ then becomes
\begin{equation} 
\Delta m = -2.5 \log_{10}\mu \, \simeq \, 5 \log_{10}(1-\kappa)\,.
\label{eq:mukappa}
\end{equation}
The lens convergence $\kappa$ can be computed from the following integral along the line of sight:
\begin{equation} 
  \kappa=\int_{0}^{r_{s}} dr 
         \frac{r(r_{s}-r)}{r_{s}} \frac{\nabla^2\Psi}{c^2} \,,
\label{k09}
\end{equation}
where $r_{s}$ is the co-moving position of the source and the integral is along an unperturbed light geodesic. The term $\nabla^{2} \Psi$ is the Laplacian of the Newtonian potential of the perturbation in co-moving coordinates. In the present case then 
\begin{equation}
 \frac{\nabla^{2} \Psi}{c^2} 
  = \frac{4\pi G}{c^2} a^2  \; \delta \rho 
  = \frac{3}{2}\Omega_{M,0}\frac{a_{0}^{2} \, H_0^2} {c^2} 
    \frac{a_0}{a} \Big( \sum_{j=1}^{N_{H}} \varphi_j -1 \Big) \,,
\label{eq:kappa2}
\end{equation}
where we have used Eqs.~(\ref{hhh}-\ref{om1}) and (\ref{deltaeq}). The sum over the halos describes the effect of inhomogeneities, while the negative constant of unity can be understood as the demagnifying potential of empty space (the ``empty beam"). Substituting Eq.~(\ref{eq:kappa2}) back to Eq.~(\ref{k09}) we get
\begin{equation}
\kappa
\equiv \kappa_H + \kappa_E 
= \int_{0}^{r_{s}}dr \, G(r,r_{s}) \Big( \sum_{j=1}^{N_{H}}\varphi_{j} -1\Big) \,,
\label{eq:kappa3}
\end{equation}
where $\kappa_E$ is the empty beam convergence and $\kappa_H$ is the convergence caused by halos, and we have defined the auxiliary function $G(r,r_{s})$:
\begin{equation}
G(r,r_{s})=  \frac{3}{2} \Omega_{M,0} \, \frac{a_0^2 H_0^2}{c^2}
             \, \frac{r(r_s-r)}{r_s} \, \frac{a_0}{a(t(r))} \,.
\label{eq:grs}
\end{equation}
In an exactly homogeneous FLRW model the two contributions $\kappa_H$ and $\kappa_E$ in Eq.~(\ref{eq:kappa3}) cancel and there is no lensing. In our setup, as we will see in detail in the next section, the two contributions cancel also if we can take the angular average over all sky. In other words, given a number of observations $N_{O}$, the GBS is obtained in the limit $N_{O} \rightarrow \infty$, if no lines of sight are obscured. When both these conditions are met, the entire volume of the space is seen and the deduced sum $\sum \varphi_{j}$ averages to unity. However, it is evident that a limited size of the data set (small $N_O$) can cause a {\em probabilistic} deviation of the PBS from the GBS, whereas selection effects can lead to {\em systematic} deviation of the PBS from the GBS. One can imagine several qualitative selection effects that could cause an apparent demagnification of the data sample, such as rejection of ``outliers", intergalactic extinction or foreground light contamination. The simplest way to estimate such effect is to assume that some fraction of the halo mass is ``hidden" from the observations.

Note also that for given parameters $(R,n_{c})$, the density contrast scales linearly with $\Omega_{M}$ so that a flat dark-energy--dominated model will have smaller lensing effects than does a flat matter-dominated EdS model.

\subsubsection*{The accuracy of the weak lensing approximation}
\label{sec:accuracy}

\begin{figure}
\begin{center}
\includegraphics[width=8.2 cm]{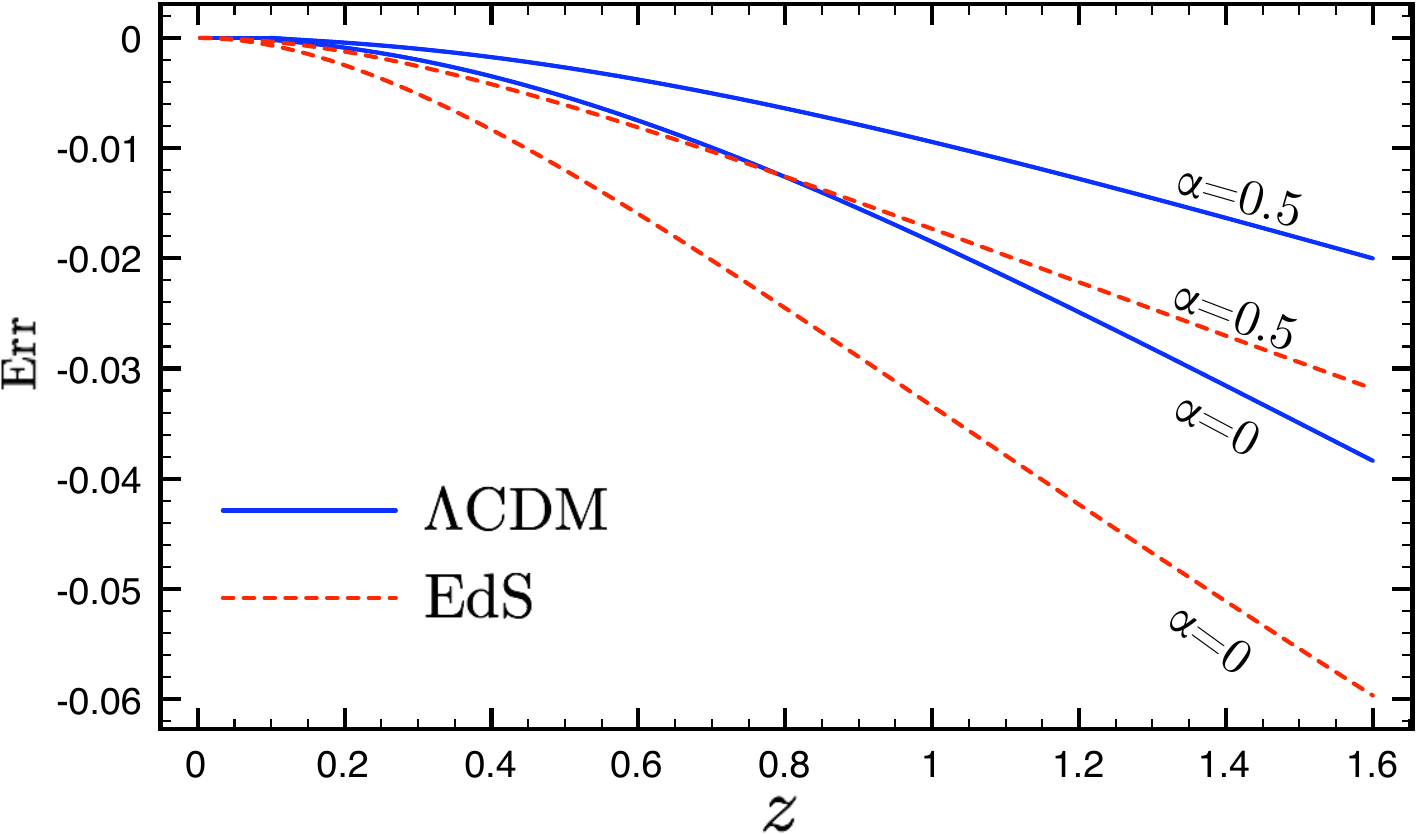}
\caption{The relative error Err of the weak-lensing calculation defined in Eq.~(\ref{errr}) for the empty beam ($\alpha=0$) and for a half filled beam ($\alpha = 0.5$)
for both the EdS and the $\Lambda$CDM model.}
\label{check}
\end{center}
\end{figure}

Above we defined a cumulative expression for the weak lensing convergence $\kappa$, whereas the observable flux is actually multiplicatively magnified by the subsequent lenses. In practice this works well when the cumulative magnification is small. That is, when $\kappa, \, |\gamma_{s}| \ll 1$ we have
\begin{equation}
\mu \simeq 1 + 2\, \kappa + 3 \, \kappa^{2} +|\gamma_{s}|^{2}+... \simeq 1 + 2\, \kappa
\end{equation}
and $\langle \kappa \rangle =0$, $\langle \mu \rangle =1$ and $\langle \Delta m \rangle=0$ are all equivalent. The accuracy of the weak lensing scheme can be tested quantitatively by a comparision with exact solutions. First, it was shown by Ref.~\cite{Vanderveld:2008vi} that the weak-lensing approximation reproduces the full general relativistic results of Ref.~\cite{Marra:2007pm} to a $\sim 5$\% accuracy  along a line of sight in an inhomogenous swiss-cheese model\footnote{The authors of Ref.~\cite{Vanderveld:2008vi} found $\Delta m \simeq 0.346$ at redshift $z\simeq 1.86$ through a particular grid of LTB-bubbles. They claimed only an accuracy of $\sim 10$\% probably because, unlike us, they did not have the exact results of Ref.~\cite{Marra:2007pm}.}, where the difference comes from neglecting higher order terms in the expression for $\kappa$ in Eq.~(\ref{k09}).

Let us next consider the solutions for empty or partially filled beams with a filling factor $\alpha\equiv \rho_{\rm beam}/\bar\rho$. Physically this corresponds to the case where a fraction $1-\alpha$ of the mass in the halos is for some reason hidden from the observations and the remaining mass fraction $\alpha$ is observed as a smooth distribution. 
Let us define the relative error between the weak lensing approximation and the exact result for a partially filled beam:
\begin{equation} \label{errr}
{\rm Err}= \frac{\Delta m^{wl}_{E,\alpha}-\Delta m^{ex}_{E,\alpha}}{\Delta m^{ex}_{E,\alpha}} \,.
\end{equation}
In Fig.~\ref{check} we show Err in the case $\alpha=0$ and $\alpha=0.5$ for the EdS and the $\Lambda$CDM models. The convergence factors for the $\Lambda$CDM model were computed integrating numerically the results of Ref.~\cite{Kantowski:1998ju}.  Analytic expressions for the exact and weak lensing convergences $\kappa_{E,\alpha}$ in the EdS model are shown in Appendix \ref{weds}. Clearly the weak lensing and exact results agree to within a few percent over the interesting redshift range. In this paper we are concerned not with all possible magnifications but with the most probable ones. Since these are bound in magnitude by the empty beam convergence, the results of Fig.~\ref{check} then suggest that the weak lensing approximation should be very good for our purposes.

\section{Probabilistic study of weak lensing} 
\label{wlcal}

Our next task is to compute the probability distribution function (PDF) and the most likely value of the lens convergence $\kappa$ along arbitrary photon geodesics as a function of our model parameters $(\Omega_{M,0}, H_0, R, n_c, N_O)$. Since we already know how to calculate $\kappa_E$ it remains to evaluate the halo-induced part $\kappa_H $ in Eq.~(\ref{eq:kappa3}):
\begin{equation}
\kappa_H (z_s) = \int_0^{r_s}dr \, G(r,r_s) \sum_{j=1}^{N_{H}} \varphi(|r-r_j|) \,,
\label{eq:kappa3b}
\end{equation}
where $r_s = r(z_s)$. We wish to obtain a probabilistic prediction for this quantity along a random line of sight to a source located at $r_s$, in a universe with randomly distributed halos. One way to solve the problem would be to construct explicitly a large enough co-moving volume with halos at fixed random locations $r_i$, and then compute $\kappa_H $ along randomly selected directions in this space. Alternatively, one can start from a fixed geodesic and construct a random distribution of halos along that geodesic. We choose the latter approach, because it allows us to find fast numerical and analytical solutions for the key quantities we are interested in. 

\begin{figure}
\begin{center}
\includegraphics[width=8.0 cm]{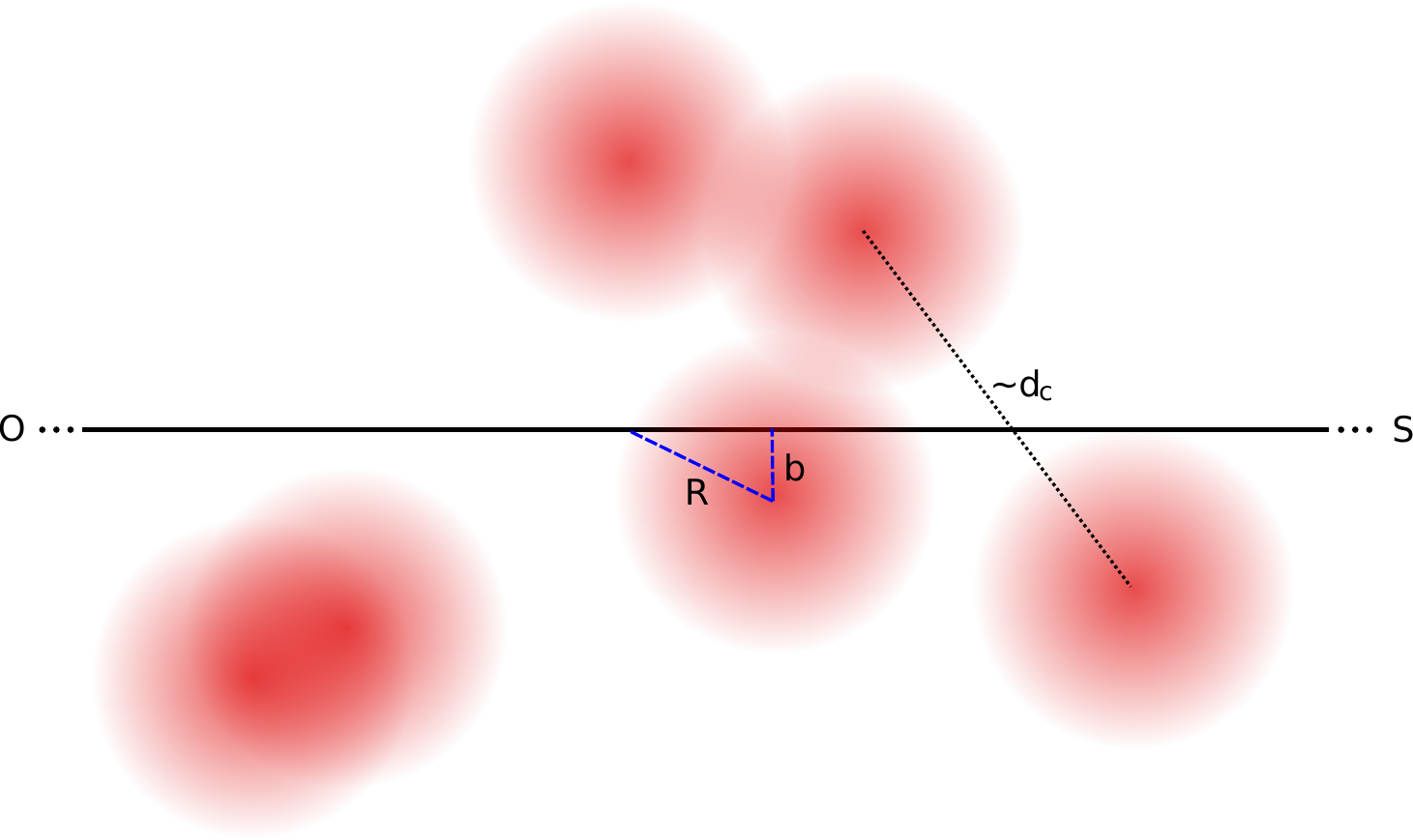}
\caption{Shown is a co-moving segment of a photon geodesic between an observer at $O$ and a source at $S$. The shaded disks represent halos of radius $R$.}
\label{sketch}
\end{center}
\end{figure}

First consider a particular {\em realization} $\theta$ of the integral in Eq.~(\ref{eq:kappa3b}), that is, a particular configuration of the halos and a particular line of sight. Because of the finite size of the halos, only the $N_\theta$ halos with impact parameters $b_j < R$ from the geodesic contribute to the sum (see the sketch in Fig.~\ref{sketch}).
Moreover, because the halos are small compared with the horizon scale, the function $G$ is to a good approximation a constant $G(r,r_s) \approx G(r_j,r_s)$ within each halo. Similarly, the time dependence of the halo profile will be weak, $\varphi(x,t) \approx \varphi(x,t_j)$.
After a little algebra one then finds:
\begin{eqnarray}
\kappa_H (\theta, z_s) 
&\simeq& \sum_{j=1}^{N_\theta}G(r_j,r_s)\int_{b_j}^{R_j} \frac{2 \,   x \, {\rm d}x}{\sqrt{x^2-b_j^2}} \varphi(x,t_j)
\nonumber \\
&\equiv&  \sum_{j=1}^{N_\theta}G(r_j,r_s)\,\Gamma(b_j,t_j) \,,
\label{eq:realization1}
\end{eqnarray}
where $R_j\equiv R(t_j)$ and $t_j = t(r_j)$.
Our sample realization is now characterized by a set of values $\{r_j,b_j\}$, where $j$ runs from 1 to $N_\theta$.

%
%

Next divide the geodesic in $N_S$ subintervals of (possibly variable) length $\Delta r_i \gg R$, such that $G(r,r_s)$ can still be taken a constant within each interval. In practice one can always construct such a division with $r_i \gg \Delta r_i \gg R$. When these conditions are met, all halos in a given bin $r_j \in [r_i,r_{i+1}]$ can be associated with the same distance $\tilde r_i$ to the center of the bin, and (\ref{eq:realization1}) becomes
\begin{equation}
\kappa_H (\theta, z_s) \simeq \sum_{i=1}^{N_S}G(\tilde r_i,r_s) \sum_{l=1}^{N_i} \Gamma(b_l,t_i) \,,
\label{eq:realization2}
\end{equation}
where $\sum_{i=1}^{N_S} N_i = N_\theta$. Now divide also the impact parameter into $N_R$ bins of width $\Delta b_m$, such that we can take $\Gamma (b,t)$ a constant within each of these bins. In this way we can rewrite (\ref{eq:realization2}) as
\begin{equation}
\kappa_H (\theta, z_s) \simeq 
\sum_{i=1}^{N_S} \sum_{m=1}^{N_R} k^\theta_{im}\,  G(\tilde r_i,r_s) \Gamma (\tilde b_m, t(\tilde r_i)) \,,
\label{eq:realization3}
\end{equation}
where $\tilde b_m$ is the average $b$ within the bin $[b_m,b_{m+1}]$ and
\begin{equation}
\sum_{i=1}^{N_S}\sum_{m=1}^{N_R} k^\theta_{im} = N_\theta \,.
\end{equation}
Now observe that $G(r,r_s)$ and $\Gamma (b,t)$ are universal functions for arbitrary values of $r$ and $b$, and so all information specific to the particular realization $\theta$ in the equation (\ref{eq:realization3}) is given by the set of integers $\{k^\theta_{im}\}$ giving the number of halos within the distance and impact parameter bins characterized by the positions $\{\tilde r_i,\tilde b_m\}$ and sizes $\{\Delta r_i,\Delta b_m\}$. 

Equation (\ref{eq:realization3}) is the starting point of our analysis, because it can be easily turned into a probabilistic quantity. Instead of considering a set of realizations $\theta$ along arbitrary lines of sight through a pre-created model universe, we can {\em define a statistical distribution of convergences through} Eq.~(\ref{eq:realization3}) because, as we have showed in Section~\ref{sec:inohomogenities1}, the integers $k^\theta_{im}$ are distributed as Poisson random variables of parameter $\Delta N_{im}$:
\begin{equation}
P_{k_{im}} = \frac{(\Delta N_{im})^{k_{im}}}{k_{im}!} e^{-\Delta N_{im}} ,
\label{eq:poisson-i}
\end{equation}
where $\Delta N_{im}$ is the expected number of halos in the bin volume $\Delta V_{im}$:
\begin{equation}
\Delta N_{im} = n_c \Delta V_{im} = n_c \, 2\pi b_m\Delta b_m \Delta r_i\,.
\end{equation}
Equivalently, we can interpret $\Delta N_{im}$ as the mean number of collisions of a photon with halos within the $i$'th $r$-subinterval and within the $m$'th impact parameter bin.
Physically this statistical distribution of convergences is equivalent to our original set of realizations $\theta$ averaged also over the position of the observer. Thus the statistical  model explicitly incorporates the Copernican Principle.

The basic quantity in our probabilistic treatment is a {\em configuration} of random integers $\{k_{im}\}$. The convergence corresponding to a given configuration is given by an equation analogous to Eq.~(\ref{eq:realization3}):
\begin{equation}
\kappa_H (\{k_{im}\},z_s) = \sum_{i=1}^{N_S} \sum_{m=1}^{N_R} 
k_{im} \, \kappa_{1im}(z_s) \,,
\label{eq:kappaMfinal}
\end{equation}
where $\kappa_{1im}(z_s) \equiv G(\tilde r_i,r_s)\Gamma (\tilde b_m,t_i)$ is the convergence due to a single halo at distance $\tilde r_i$ and impact parameter $\tilde b_m$. The probability that such configuration occurrs is just
\begin{equation}
P_{\{k_{im}\}} = \prod_{i=1}^{N_S} \prod_{m=0}^{N_R} P_{k_{im}} 
           \equiv \prod_{im} P_{k_{im}}     \,.
\label{eq:Prob1}
\end{equation}
It is easy to see that this probability is normalized to one:
\begin{equation}
\sum_{\{k_{im}\}} P_{\{k_{im}\}} = \prod_{im} \sum_{k_{im}=1}^\infty P_{k_{im}} 
= 1    \,.
\end{equation}
The expectation value for the convergence is given by the probability weighted sum over all possible $\{k_{im}\}$ configurations. Using the above results it becomes:
\begin{eqnarray}
\langle \kappa_H (z_s)\rangle &\equiv& 
\sum_{\{k_{im}\}} P_{\{k_{im}\}} \kappa_H (\{k_{im}\},z_s) 
\nonumber \\ 
&=&
\sum_{im} \, \langle k_{im}\rangle  \kappa_{1im}(z_s) 
\nonumber \\ 
\nonumber \\ 
&=&
\sum_{im} \, \Delta N_{im} \kappa_{1im}(z_s) 
\nonumber \\ 
&=&
\sum_{i=1}^{N_S} \, G(r_i,r_s) \Delta r_i = -\kappa_E \,.
\label{eq:kappa3d}
\end{eqnarray}
That is, the total expected convergence $\langle \kappa \rangle = \langle \kappa_H \rangle + \kappa_E$ vanishes bringing back the GBS result, consistent with photon conservation in weak lensing.  In going from the third to the last line we used the identity
\begin{eqnarray}
&&\sum_{m=1}^{N_R} 2\pi b_m \Delta b_m \, n_c \Gamma (\tilde b_m,t_i) 
\nonumber \\
&& \phantom{hannatyt} = 2\pi n_c \int_{0}^{R} {\rm d}b \, b\Gamma (b,t) = 1 \,,
\label{eq:identity}
\end{eqnarray}
which, when calculated in the continuum limit, holds true for any functional form of the halo profile $\varphi$ with the normalization of Eq.~(\ref{eq:normof-f}) as shown in Appendix \ref{identity}. Finally note that, using the identity  (\ref{eq:identity}), we can write the total convergence for an observation corresponding to a configuration $\{k_{im}\}$ simply as:
\begin{equation}
\kappa(\{k_{im}\},z_s) = \sum_{i=1}^{N_S} \sum_{m=1}^{N_R}  \,\kappa_{1im}(z_s) \Big(k_{im} -\Delta N_{im} \Big),
\label{eq:fullkappa}
\end{equation}
where $k_{im}$'s are random variables drawn from the Poisson distribution (\ref{eq:poisson-i}): $k_{im} \equiv k_{im}[\Delta N_{im}]$. Eq.~(\ref{eq:fullkappa}) makes explicit that the expected convergence vanishes: the expected value of each term of the summation is indeed zero.

Note that in the analysis above, we did not give any quantitative criterion for binning the variables $r$ and $b$. The fact that both $\Delta r_i$ and $\Delta b_i$ formally vanish in the final expression in Eq.~(\ref{eq:kappa3d}) already suggests that the exact way the binning is done is not important. This is indeed so, and one can even show formally that the binning is not important as long as it is sufficiently fine to give a good approximation for the functions $G(r,r_s)$ and $\Gamma (b,t)$. One can also easily test the effect of binning directly by comparing the final results obtained with different size grids. In practice we found that a grid with $10-20$ points in each variable already gives very accurate results.  

The final convergence PDF can be formally written as
\begin{equation}
P_{\rm wl}(\kappa, z_s) = \sum_{\{k_{im}\}} P_{\{k_{im}\}} 
                 \delta(\kappa - \kappa(\{k_{im}\},z_s)) \,.
\label{eq:PDF1}
\end{equation}
In the continuum limit this becomes formally a functional integral over random integer-valued functions $k(r,b)$. It is clear that the most likely configuration which maximises the probability function $P_{\{k_{im}\}}$ corresponds to the mode: $k_{im} \rightarrow \lfloor \Delta N_{im}\rfloor$. Moreover, for large $\Delta N_{im}$ this most likely configuration approaches the mean $k_{im} \rightarrow \Delta N_{im}$. When this is the case the expectation value for total convergence vanishes even for a single observation. This signals the fact that the PDF then approaches a gaussian with a vanishing skewness.

One can also create the probability distribution directly for the shift in distance modulus $\Delta m$:
\begin{equation}
P_{\rm wl}(\Delta m, z_s) = 
\sum_{\{k_{im}\}} P_{\{k_{im}\}} \,\delta(\Delta m - \Delta m(\{k_{im}\},z_s))\,,
\label{eq:PDF2a}
\end{equation}
where one uses equation (\ref{eq:mukappa}) to compute $\Delta m(\{k_{im}\},z_s)$ from the converegence.
Similarly, one can also define the PDF form the magnification $\mu$.

It is straightforward to compute the distributions (\ref{eq:PDF1}) and (\ref{eq:PDF2a})  through a numerical simulation, simply by creating a sufficiently large set of random configurations $\{k_{im}\}$ drawn from the probability distribution $P_{\{k_{im}\}}$. Note that the distributions (\ref{eq:PDF1}) and (\ref{eq:PDF2a}) are formally discrete but very dense sets of distributions whose integral over $\kappa$ or $\Delta m$ is normalized to unity. So the quantity that one actually computes through a simulation is: 
\begin{equation}
P_{\rm wl}(\Delta m_l, z_s) = 
\int_{\Delta m_l-\Delta_{bin}/2}^{\Delta m_l+\Delta_{bin}/2} {\rm d} yÊ\,
P_{\rm wl}(y, z_s) \,,
\label{eq:PDF2}
\end{equation}
where $\Delta_{bin}$ are some suitably chosen bin widths. The PDF (\ref{eq:PDF2}) is one of the main results of this paper. We wish to stress the simplicity of this result. In order to specify the model completely one needs only Eqs.~(\ref{eq:grs}) and (\ref{eq:realization1}) for the functions $G(r,r_s)$ and $\Gamma (b,t)$ respectively. After this, $P(\Delta m_l, z_s)$ is found from the random sample of magnifications $\Delta m$
computed through equations (\ref{eq:fullkappa}) and (\ref{eq:mukappa}). 

\subsection{Selection effects} 
\label{sec:selection}

As was discussed in Section ~\ref{sec:cumulative}, several selection effects might bias the convergence distribution, favouring overall demagnification. It is straightforward to generalize our probabilistic approach to include many such effects. For example, all effects that would lead to a rejection of a potential SN observation can easily be described by an additional {\em survival probability function}. That is, we replace
\begin{equation} 
P_{\{k_{im}\}} \rightarrow P^{\rm eff}_{\{k_{im}\}} \equiv K \, P^{\rm sur}_{\{k_{im}\}}P_{\{k_{im}\}}
\label{eq:survival}
\end{equation}
in our master formula for the magnification PDF (\ref{eq:PDF2}). Here $K$ is a normalization constant that makes sure that the final PDF is normalized to unity. The most generic form of the survival probability function is
\begin{equation}
P^{\rm sur}_{\{k_{im}\}} = \prod_{im} (P^{\rm sur}_{im})^{k_{im}} \,.
\label{eq:Pim}
\end{equation}
This allows the surival function depend on the arbitrary local properties along the photon geodesic. Given the form (\ref{eq:Pim}) the proper normalization is easily seen to be:
\begin{equation} 
P^{\rm eff}_{\{k_{im}\}} = \prod_{im} 
   \frac{(\Delta N^{\rm eff}_{im})^{k_{im}}}{k_{im}!} 
         e^{-\Delta N^{\rm eff}_{im}} \,,
\label{eq:survival2}
\end{equation}
where
\begin{equation}
\Delta N^{\rm eff}_{im} = P^{\rm sur}_{im} \Delta N_{im} \,.
\label{eq:deltaNeff}
\end{equation} 
The correct expression for the convergence is still given by Eq.~(\ref{eq:fullkappa}), with the important difference that the random integers $k_{im}$ are now drawn from the Poisson distribution Eq.~(\ref{eq:survival2}) with the effective expected number of halos: $k_{im} \equiv k_{im}[\Delta N^{\rm eff}_{im}]$. Now $\langle k_{im}\rangle = \Delta N^{\rm eff}_{im}$ so that the average convergence over a large number of observations becomes
\begin{equation}
\langle \kappa(z_s) \rangle = \sum_{im} \kappa_{1im}(z_s) (P^{\rm sur}_{im} - 1)\Delta N_{im} \,.
\label{averagekappa}
\end{equation} 
As expected, the selection biases survive in the data even after averaging over many observations. Note that in the above analysis we implicitly assumed that the survival probability for a light ray going through empty space equals unity. 

The actual form of the survival function $P^{\rm sur}_{im}$ depends on detailed input both from the astrophysical properties of the intervening matter distributions (estimate for extinction, foreground light contamination, etc.) and from the observational apparatus ({\em e.g.}~detection efficiency of SNe-triggering telescopes). A quantitative analysis of these issues is beyond the scope of the present paper, and we merely show how the method can be applied to the simple selection effect discussed in Section~\ref{sec:accuracy}. To see this, rewrite Eq.~(\ref{averagekappa}) as 
\begin{equation}
\langle \kappa(z_s) \rangle = \sum_{i=1}^{N_S} (\alpha_i-1) G(\tilde r_i,r_s) \Delta r_i \,,
\label{averagekappa2}
\end{equation} 
where 
\begin{equation}
\alpha_i \equiv n_c \sum_m 2\pi b_m \Delta b_m \, \Gamma(\tilde b_m,t_i) \,  P^{\rm sur}_{im}
\label{eq:alphai}
\end{equation}
is the fraction of the mass in the halos that is accessible for observations at $r_i$. Note that a constant $P^{\rm sur}_{im}=\alpha_i$, thanks to the identity  (\ref{eq:identity}), is a special solution to Eq.~(\ref{eq:alphai}). If $P^{\rm sur}_{im}$ did not depend on $r_i$, then this further reduces to
\begin{equation}
\langle \kappa \rangle = (1-\alpha)\, \kappa_E
\label{kalpha}
\end{equation}
with $\alpha_i =\alpha$ for all $i$.
As outlined in Section~\ref{sec:accuracy} and showed in detail in Appendix \ref{weds}, $\alpha$ can be interpreted as the filling factor $\alpha\equiv \rho_{\rm beam}/\bar\rho$ of a partial filled beam.
We see therefore that the results of Ref.~\cite{Kantowski:1969} can be obtained as a limiting case of our approach.

Other types of selection biases, such as due to an error in the estimate for reddening, would not affect the probability distributions, but the evaluation of the effective magnification factor itself.  Also these corrections could depend on the mass distribution along the photon path, and so, to be as general as possible, one should replace 
\begin{eqnarray}
\Delta m &\rightarrow& 
        \Delta m(\{k_{im}\},z_s) + \delta_{\rm s} \Delta m(\{k_{im}\},z_s)
\nonumber \\
&\equiv& \Delta m_{\rm eff}(\{k_{im}\},z_s)
\label{eq:deltameff}
\end{eqnarray}
in the integrand of Eq.~(\ref{eq:PDF2}).
Again, a quantitative estimate of the size of the bias factor $\delta_{\rm s}\Delta m$ is beyond the scope of this paper.

\subsection{Sources with intrinsic luminosity dispersion} 
\label{sec:initial}

Our results can be easily generalized for an arbitrary initial flux or magnitude spectrum of an imperfect standard candle. If we know that the source magnitudes can be described by a function $P_{\rm in}(\Delta m_0)$, then the observed magnification PDF is obtained by the convolution of the initial PDF and the weak lensing PDF
\begin{eqnarray}
&&
P(\Delta m,z_s) 
= \int {\rm d} y  \, P_{\rm wl}(y)P_{\rm in}(\Delta m - y) \nonumber\\
&& \phantom{Han}
= \sum_{\{k_{im}\}} P_{\{k_{im}\}}  P_{\rm in}(\Delta m - \Delta m(\{k_{im}\},z_s)) . \phantom{han}
\end{eqnarray}
Alternative expressions where one or both distributions are replaced by the probability distribution in the flux are trivially obtained by a simple change of variables.

\subsection{PDF for a sample of $N_{O}$ observations}
\label{sec:observations}

The effective distribution for a binned sample of $N_O$ supernova observations can formally be expressed as an iterative convolution~\cite{Holz:2004xx}:
\begin{eqnarray} \label{convico}
P_{N_O}(z_s,\Delta m) &=& N_O \int {\rm d}y P_{N_O-1}(z_s,y) \times
\nonumber \\
&& \times \, P_1(N_O\Delta m-(N_O-1)y).
\end{eqnarray}
Here $P_1$ is the normalized fundamental PDF that may include selection effects and a convolution over the initial distribution. Alternatively, one can compute $P_{N_O}$ directly from the fundamental PDF by creating a large number of sets of $N_O$ random realizations from it, and creating a normalized distribution for the average $\Delta m$ within these sets.

A third way is to create the $P_{N_O}$ distribution directly in the initial simulation, bypassing the calculation of the fundamental $P_{1}$ altogether. This reduces computational time and improves accuracy, of course with the price of limiting the amount of information available. If we label the configurations within a given sample of observations by $s$, the mean convergence after $N_O$ observations is:
\begin{eqnarray} 
\kappa_{N_O}(\{k_{im}\})&=& {1 \over N_{O}} \sum_{s=1}^{N_{O}} \kappa(\{k_{im}\}_{s}) \nonumber \\
&=&\sum_{i=1}^{N_S} \sum_{m=1}^{N_R}  \kappa_{1im} \big({ \sum_{s=1}^{N_{O}}  k_{im, s} \over N_{O}} -\Delta N_{im} \big) \nonumber \\
&=& \sum_{i=1}^{N_S} \sum_{m=1}^{N_R}  \kappa_{1im} \big( {k_{im, N_{O}} \over N_{O}} -\Delta N_{im} \big) \, ,
\label{equation:kappa3}
\end{eqnarray}
where, given that each of the $N_O$ observations are independent, we have used the fact that independent Poisson variables with the same weight sum exactly into a Poisson variable of parameter given by the sum of the individual parameters: $k_{im, N_{O}} \equiv k_{im}[N_O\Delta N_{im}]$.  $P_{N_O}$ is then given by Eq.~(\ref{eq:PDF2a}) where the magnification for a given configuration $\{k_{im}\}$ is computed from the convergence of Eq.~(\ref{equation:kappa3}).

We followed this last approach when creating the illustrations relevant for the Union Catalog and JDEM data in Section~\ref{sec:results}.
However, note that including the selection effects in general does not commute with taking the average over the observations. In other words, if the $N_{O}$ measurements are correlated then we cannot use Eq.~(\ref{equation:kappa3}), but we have to start from the fundamental PDF. On the other hand, Eq.~(\ref{equation:kappa3}) displays {\it explicitly} the effect of the size of the data sample: even when $\kappa$ has a skewed PDF and a nonzero mode for $N_{O}=1$, for large $N_{O}$ the distribution approaches a gaussian and eventually converges to a $\delta$-function at zero convergence, as it is clear from the properties of Poisson variables discussed in Section \ref{sec:inohomogenities1}.

\subsection{Arbitrary mass-distribution of halos} 
\label{sec:fm}

We have so far expressed all our derivations assuming there is only one halo type. At this point it is straightforward to generalize all our formulas to the case where one has a continuous distribution of halos. Let us assume that the halo mass distribution is given by some dimensionless function $f(M,z)$, which is related to the co-moving number density $n(M,z)$ through the standard relation
\begin{equation}
n(M,z) \equiv \frac{\bar{\rho}_{M,0} \, a_{0}^{3}}{M}f(M,z) \,.
\label{eq:fdef}
\end{equation}
We also assume that $f(M,z)$ is normalized to one, say:
\begin{equation}
\int {\rm d}M_{15}f(M_{15},z) \equiv 1 \,,
\label{eq:fdef2}
\end{equation}
where we have introduced the usual dimensionless mass parameter $M_{15} \equiv M/(10^{15}h^{-1}M_\odot)$. That is, our definition for $f(M,z)$ is analogous to the usual Press-Schecter~\cite{Press:1973iz} and the Sheth-Tormen~\cite{Sheth:1999mn} mass functions. However, our $f(M,z)$ is not expected to be {\em quantitatively} similar to the PS- and ST-functions, since the latter are designed to model the density of virialized systems, while we are also interested in effectively describing other large-scale structures (see the next Section \ref{sec:cdh}).

The first step to extend our treatment is to generalize the normalization of the halo profile $\varphi$:
\begin{equation}
4 \pi \int_{0}^{R(t,M)} \varphi(b,t,M) \, b^{2} db  =  {f(M,t) \over n(M,t)} \, ,
\end{equation}
where instead of the redshift $z$ we have used the time along the geodesic, $f(M,t)=f(M,z(t))$.
The halo mass is now given by Eq.~(\ref{eq:fdef}) and the definition of $\Gamma(b,M,t)$ follows as in Eq.~(\ref{eq:realization1}):
\begin{equation}
\Gamma(b,t,M) = \int_{b}^{R(t,M)} \frac{2 \,   x \, {\rm d}x}{\sqrt{x^2-b^2}} \, \varphi(x,t,M) \, .
\end{equation}
Finally we discretize the function $f(M,z)$ in the same way we discretized the impact parameter and the co-moving distance and the final result is:
\begin{eqnarray}
\kappa(\{k_{imn}\},z_s) = \sum_{i=1}^{N_S} \sum_{m=1}^{N_R} \sum_{n=1}^{N_M} \,\kappa_{1imn}(z_s) \times \nonumber \\
 \times  \Big(k_{imn}-\Delta N_{imn} \Big) ,
\label{eq:fullkappac}
\end{eqnarray}
where  $\kappa_{1imn}(z_s) \equiv G(\tilde r_i,r_s)\Gamma (\tilde b_m,t_i,\tilde M_n)$ and $\tilde M_n$ is the mean $M$ in the $n$'th mass bin.
The parameter of the Poisson variable $k_{imn} \equiv k_{imn}[\Delta N_{imn}]$ now is
\begin{equation}
\Delta N_{imn} = n(M_{n},t_{i})\Delta M_n \; 2\pi b_m\Delta b_m \; \Delta r_i\,.
\label{eq:Nimn}
\end{equation}
The probability distribution for the shift in distance modulus $\Delta m$ now becomes
\begin{eqnarray}
P_{\rm wl}(\Delta m, z_s) &=& 
\sum_{\{k_{imn}\}} P_{\{k_{imn}\}} \times
\nonumber \\ &&
\times \,\delta(\Delta m - \Delta m(\{k_{imn}\},z_s))\,.
\label{eq:PDF2b}
\end{eqnarray}
One can similarly extend the systematic effects to be dependent on the mass, such that the survival probability $P_{im} \rightarrow P_{imn}$, whereby $\Delta N^{\rm eff}_{im} \rightarrow \Delta N^{\rm eff}_{imn}$ in Eq.~(\ref{eq:deltaNeff}), and also $\Delta m \rightarrow \Delta m_{\rm eff}(\{k_{imn}\},z_s)$ in Eq.~(\ref{eq:deltameff}). Finally, Eq.~(\ref{eq:PDF2}) formally retains its present form. Eqs.~(\ref{eq:fullkappac}-\ref{eq:PDF2b}) can be used to compute all the interesting quantities for an arbitrary halo mass function $f(M,z)$.
However, for simplicty, we will from now on concentrate only on single-mass functions in this paper.
See the model of Ref.~\cite{Kainulainen:2009sx} for an example with two types of halos.

\subsection{Discussion} 
\label{sec:cdh}

Our stochastic analysis of lensing presented above is based on using an unclustered Poisson distribution of halos. Let us now discuss the applicability of this assumption and the possible ways to improve the clustering algorithm.

Ref.~\cite{Holz:1997ic} states that the lensing properties of a universe made of point masses are independent of their masses and clustering.
The statement was confirmed numerically in two steps.
First, it was shown that random distributions of point masses of $M_{\odot},\, 10^{12}M_{\odot} \textrm{ and } 10^{13} M_{\odot}$ share the same lensing properties. Second, Ref.~\cite{Holz:1997ic} compared a universe made of point mass galaxies with a universe made of uniform density galaxies composed of point mass stars. The authors concluded with their Fig.~10 that, for an intergalactic separation of $2$ Mpc (which corresponds to a mass of $\sim10^{12}M_{\odot}$) and a radius for the uniform density galaxies of $200$ kpc, the above two universes have the same lensing properties.

These results, however, do not apply for extended halos, and anyway do not prove that clustering can be neglected for objects with masses larger than $\sim10^{13} M_{\odot}$. On the contrary, we will show in Section \ref{sec:results} that the lensing PDF changes significantly when one considers halos of the scale of large superclusters of galaxies (compare Fig.~\ref{PDFs} with Fig.~\ref{PDFs3}). Let us stress again, as we pointed out in Section \ref{deschalo}, that by ``halo'' we do not mean only gravitationally bound systems, but (actually the surface mass projection of) any unvirialized sufficiently localized large-scale structures. The idea is that an unclustered Poisson distribution of large-scale structures can give a qualitative estimate of the lensing effects produced by the actual clustering of galaxies into filaments and walls. This idea was carried out further in the companion work of Ref.~\cite{Kainulainen:2009sx} where structures of average separation of $d_c = 100 \, h^{-1}$ Mpc (which corresponds to a mass of $6 \cdot 10^{17} h^{-1} M_{\odot}$) and radius of $R_p = 10 \, h^{-1}$ Mpc were considered.

Of course randomly placed spherical structures is but a crude approximation for the actual 3-dimensional web-like structure of clusters and filaments. However,  all that matters for weak lensing are the projected 2D-surface mass densities over a series of spatial slices from the observer to the source. For these projections the difference between a realistic 3D-structure and the crude meatball model is less distinctive. Nevertheless, it would be desirable to improve the clustering algorighm, and we can imagine several ways to do it within our approach. First, our most general result in the form of Eq.~(\ref{eq:fullkappac}) allows us to deal with a generic halo mass distribution function $f(M,z)$. In the most conservative approach one would identify $f$ with some known clustering function, such as the Press-Sechter distribution. However, to incorporate also the noncoherent structures one could use instead an effective form, say $f(M,z) \sim N_\xi f_{PS}(\xi M,z)$, where $N_\xi$ is a normalization factor.  For $\xi<1$ such form would in a simple way model the merging of smaller halos into larger systems of (unvirialized) clusters. Another, and probably a more accurate method to describe very large scale clustering would be to use a master probability  distribution function to modulate the average matter background density on the scale ${\cal O}(100)$ Mpc, along the lines of sight, together with a normal local clustering fuction $f(M,z,\delta \rho(z))$, where $\delta\rho$ is the modulation around the GBS density. This method also allows an easy way to implement a local void around the observer. Yet another way to improve the clustering algorithm would be to introduce non-spherical structures, such as cylinders to better describe filaments and walls. However, we suspect that while such extension would apparently help gaining a more realistic 3D-distribution, it would not be crucial for the modelling of the surface density projections.

\section{Analytic results for $P_{\rm wl}(\Delta m, z_s)$}
\label{sec:analytic}

In this chapter we will derive an analytic approximation for the magnification PDF defined in the previous Section, in particular we will focus on mode and variance. 
A reader not interested in details can skip to the final results presented in Eqs.~(\ref{kappaka}), (\ref{eq:finalbias}) and (\ref{eq:finaldispersion}); the validity of these expressions will be tested numerically in the Section~\ref{sec:results}.
The procedure will consist of two steps of resummation over the variables $b_m$ and $r_i$. Let us first introduce the central idea for these summations using a generic example depending on one set of variables.

\subsection{Linear combination of Poisson random variables}

Consider a collection of independent Poisson random variables $k_{i}$ with parameters $\lambda_{i}$, $k_i[\lambda_{i}]$.
As we have seen in the derivation of Eq.~(\ref{equation:kappa3}), the sum  $k = \sum_{i}k_{i}$ is then a Poisson random variable with parameter $\lambda = \sum_{i}\lambda_{i}$.
However, we wish to study the distribution of the quantity
\begin{equation}
S= \sum_{i} Z_{i} \, k_i[\lambda_i] \,,
\end{equation}
where the weights $Z_{i}$ are positive real numbers. When the weights $Z_{i}$ are different, no simple exact expression exists for the probability distribution of $S$. We shall adapt the following approximative procedure:
\begin{eqnarray} 
S &=& \sum_{i} Z_{i} \, k_i[\lambda_i]
   = \bar{Z} \sum_{i} z_{i} \, k_i[\lambda_i] 
\nonumber\\
  &\approx& \bar{Z} \sum_{i}  k_i[z_i \lambda_i]
   = \bar{Z} \,  k[\sum_{i} z_{i} \lambda_{i}] 
   \nonumber\\
  &=& \bar{Z} \,  k[\bar \lambda]
  \equiv \tilde{S} \,,
\label{pro1}
\end{eqnarray}
where we have defined $\bar{Z}=\max\{Z_{i}\}$ and $z_{i} = Z_{i} / \bar{Z}$. The new variable $\tilde{S}$ is Poisson distributed with the parameter
$\bar \lambda \equiv \sum_{i} z_{i} \lambda_{i}$. The idea of the approximation made in the third step of Eq.~(\ref{pro1}) is to use the normalized weight distribution $z_i$ to define a new set of random variables $z_i\lambda_i$ that favour the terms that give the largest contribution to the original sum $S$. In this way, the PDF for $\tilde S$ should provide a reasonable approximation for the mode and the skewness of the actual distribution.
It easy to show that this procedure preserves the mean and works exactly for the trivial case of constant $Z_{i}$. However, the approximation distorts the variance of the distribution:
\begin{eqnarray}
\sigma^2(S)&=& \bar{Z}^{2} \sum z^{2}_{i} \lambda_{i} \\
\sigma^2(\tilde{S})&=& \bar{Z}^{2} \sum z_{i} \lambda_{i} \, .
\end{eqnarray}
Thus the dispersion $\sigma$ for $\tilde S$ should be be corrected with:
\begin{equation}
\omega = \left( \frac{\sum_i z^2_i \lambda_i}{\sum_i z_i \lambda_i} 
         \right)^{1/2} \,.
\label{eq:dispersion}
\end{equation}
%

\subsection{Summation in $b_m$}
Let us now apply the results of the previous Section to reduce Eq.~(\ref{eq:fullkappa}). Here we have two independent indices to be accounted for, and we shall resum them sequentially starting from the summation in $b_m$. Writing only the relevant pieces we have
\begin{equation}
\sum_{m=1}^{N_R} \Gamma (b_{m}, t_{i}) \, k_{im}[\Delta N_{im}] 
\approx
\bar{\Gamma }_i  \, k_i[\Delta N_i]\,,
\label{eq:bsum}
\end{equation}
where $\bar{\Gamma }_i \equiv {\rm max}\{\Gamma (b_{m}, t_{i})\}$ and
$k_i$ is the new effective Poisson distributed random variable with an effective parameter $\Delta N_i$:
\begin{equation}
\Delta N_i = \frac{\Delta r_i}{\bar \Gamma_i} 
             \sum_{m=1}^{N_R}  2 \pi b_m \Delta b_m \, n_c \, \Gamma (b_m, t_i)
           = \frac{\Delta r_i}{\bar \Gamma_i} \,,
\label{eq:delNieq}
\end{equation}
where we again used Eq.~(\ref{eq:identity}). Inserting the approximation (\ref{eq:bsum}) in Eq.~(\ref{eq:fullkappa}) we then find
\begin{equation}
\kappa = \sum_{i=1}^{N_S} G(r_i,r_s)
    \left( \bar \Gamma_i \, k_i[\Delta N_i] - \Delta r_i \right) \,.
\label{dbr}
\end{equation}
Before resumming the $r$-variable we pause to give estimates for the quantity $\bar \Gamma$. These obviously depend on the choice of the halo density profile. First consider the uniform density halo with $\varphi_{\rm uni} = 3/(4\pi R^3 n_c)$. This implies 
\begin{equation}
\Gamma_{\rm uni}(b,R) = \int_b^{R} \frac{2 \, x \, dx}{\sqrt{x^2-b^2}} \varphi_{\rm uni}=  
\frac{3\sqrt{R^2-b^2}}{2\pi R^3 n_c}
\end{equation}
so that 
\begin{equation}
\bar \Gamma_{\rm uni} = \Gamma_{\rm uni}(0,R) = \frac{3}{2 n_c\pi R^2 }\,.
\end{equation}
Second, for a correctly normalized gaussian 
\begin{equation} 
\varphi_{\rm gau}(r)= \frac{27}{n_c \, R^3 (2 \pi)^{3/2}} \exp \left(-\frac{9 \, r^2}{2R^2} \right)
\label{gau1}
\end{equation}
one finds
\begin{equation}
\bar \Gamma_{\rm gau}=\frac{9}{2n_{c}  \, \pi  R^{2}} \,.
\end{equation}
These results suggest we parametrize our $\bar \Gamma_i$ as:
\begin{equation} 
\bar \Gamma_i^{-1} \equiv n_c  \, \pi  R_i^2 \, Q_\varphi^{2} \,,
\label{eq:barFi}
\end{equation}
where $R_i = R(t_i)$. 
That is, $\bar \Gamma_i$ correponds to an effective mean free path of a photon at $r\approx r_i$. The effective distribution parameter $\Delta N_i$ now becomes:
\begin{equation}
\Delta N_i = n_c \, \pi  R_i^2 Q_\varphi^2 \, \Delta r_i \,.
\label{eq:deltaNi}
\end{equation}
For example for the gaussian profile above it is $Q_{\varphi}=\sqrt{2}/3\simeq 0.47$.
Equations (\ref{eq:barFi}) and (\ref{eq:deltaNi}) specify the distribution (\ref{dbr}) completely in terms of a single effective parameter $Q_\varphi$, which depends on the particular density profile $\varphi$ chosen.
This method of estimating analytically $Q_\varphi$ fails for profiles that are singular at the origin, such as the sigular isothermal sphere (SIS). In these cases $Q_\varphi$ can be fitted from the numerical distribution. In any case, the essential point is that $Q_\varphi$ can be taken as a constant, and the general rule is that the more peaked the profile is, the smaller effective $Q_\varphi$ one finds. Table~\ref{Qphi} at the end of this Section shows the numerical values used in Section \ref{sec:results}.

\subsection{Summation in $r_i$}

We now repeat the resummmation approximation for Eq.~(\ref{dbr}). Again writing only the relevant term, we get
\begin{equation} 
\kappa_H = \sum_{i=1}^{N_S} H_i(r_s) \,k_i[\Delta N_i]
\approx \bar H(r_s)  k[N_{C}] \,,
\label{dr1}
\end{equation}
where $H_i(r_s) \equiv G(r_i,r_s)\bar \Gamma_i$ and again $\bar H(r_s) \equiv {\rm max}\{H_i\}$. Finally  
\begin{equation}
N_{C} \equiv \sum_{i=1}^{N_S} h_i\Delta N_i \,,
\label{eq:deltaN1}
\end{equation}
with $h_i \equiv H_i/\bar H$. Inserting the expression Eq.~(\ref{eq:delNieq}) for $\Delta N_i$ into Eq.~(\ref{eq:deltaN1})
we get
\begin{equation}
N_{C} = \frac{\bar G(r_s)}{\bar H(r_s)} \sum_{i=1}^{N_S} g_i(r_s) \Delta r_i 
\approx \frac{\bar G(r_s)}{\bar H(r_s)} \, r_s \, E \,,
\label{eq:deltaN2}
\end{equation}
where we again defined $\bar G(r_s) \equiv {\rm max}\{G(r_i,r_s)\}$ and  $g_i(r_s) \equiv G(r_i,r_s)/\bar G(r_s)$. The function $g$ depends very weakly on $r_s$ and one finds that numerically
\begin{equation}
E \approx \int_{0}^{1}dx \, g(x)  \simeq {2 \over 3} \,.
\end{equation}
Finally one can show that to a very good accuracy
\begin{equation} \label{r4}
{\bar{G}(r_{s}) \over \bar{H}(r_{s})} \simeq n_c  \, \pi  \bar R^2 Q_\varphi^2  \,,
\end{equation}
where $\bar R \equiv R(t(r_s/2))$. Inserting these results back to Eq.~(\ref{eq:deltaN1}) we finally get the estimation:
\begin{equation}
N_{C} \approx   n_{c} \cdot r_{s} \, E \cdot  \pi  \bar{R}^{2} Q_{\varphi}^{2}\, .
\label{eq:finalDeltaN}
\end{equation}
Physically $N_{C}$ corresponds to the expected number of collisions with halos in a tube of effective radius $\bar R Q_{\varphi}$ and effective length $r_{s} E$ connecting observer to source.
Our final result is an effective convergence function which is Poisson distributed with parameter $N_{C}$:
\begin{equation}
\kappa(k,z_s) = \kappa_E \left(1 - {k[N_{C}] \over N_{C}} \right) \,.
\label{eq:approxpoisson}
\end{equation}
Note that the average convergence for this distribution vanishes as it must.

\subsection{Final analytic result}

Both our exact result of Eq.~(\ref{eq:fullkappa}) and our approximate result of Eq.~(\ref{eq:approxpoisson}) are representative of the actual PDF for magnification and convergence. Often one is rather interested in the probabilistic interpretation of a set of $N_O$ identical, or sufficiently similar observations. For example, one typically introduces some binning of the data points, and one would like to know what is the most likely value and the dispersion for such an effective observable.
Repeating the derivation of the result of Eq.~(\ref{equation:kappa3}), we can extend Eq.~(\ref{eq:approxpoisson}) to a sample of  $N_O$ observation by replacing  $N_{C}$ with the total effective number of collisions:
\begin{equation}
N_T(z) \equiv N_O(z) N_{C}(z) \,,
\end{equation}
so that the convergence for a configuration $k$ is
\begin{equation}
\kappa(k,z_s) = \kappa_E \left(1 - {k[N_T] \over N_T} \right)  \,.
\label{kappaf}
\end{equation}
Finally, the approximate probability distribution function in magnitudes is:
\begin{equation}
P_{\rm wl}(\Delta m, z_{s}) = e^{- N_T} \frac{N_T^k}{k!} \, ,
\label{kappaka}
\end{equation}
where $k$ solves $\Delta m = 5 \log_{10}(1-\kappa(k, z_{s}))$.
The dispersion of Eq.~(\ref{kappaka}) has to be corrected by the factor $\omega$ to be derived below in Section \ref{sedispe}.

The distribution of Eq.~(\ref{kappaka}), similarly to the exact result of Eq.~(\ref{eq:PDF2a}), is discrete.
For the latter the chosen bin width $\Delta_{bin}$ of Eq.~(\ref{eq:PDF2}) is irrelevant because the PDF is a very dense in its domain.
This is not, however, the case for Eq.~(\ref{kappaka}) and so we have to give an explicit prescription for the binning.
In order to estimate the order of magnitude of $\Delta_{bin}$, we evaluate the convergence caused by one halo placed at half way between observer and source and hit with impact parameter $\bar{R} \, Q_{f}$:
\begin{equation}
\bar \kappa_{1} \sim G(r_{s}/2,r_{s}) \, \Gamma(\bar{R} \, Q_{f},t(r_{s}/2)) \, .
\end{equation}
The bin width will then be $\Delta_{bin} \sim  5 \log_{10}(1-\bar \kappa_{1})$.

\subsection{Mode}

The mode for the effective convergence is obviously
\begin{equation}
\kappa_{B}(z) = \kappa_{E}(z) \left(1- {\lfloor N_{T}(z) \rfloor \over N_{T}(z)} \right) \,.
\label{eq:finalformode}
\end{equation}
We denoted the mode by $\kappa_{B}(z)$ to emphasize its meaning as the {\em bias} away from the homogenous limit $\kappa = 0$. The final step in our derivation is to remove the discreteness of the probability distribution by introducing the continuum approximation $\chi(x)$ for the floor function:
\begin{equation}
{\lfloor x \rfloor \over x} \longrightarrow
\chi(x)=
\begin{cases}
 0 & 0\le x \le 1/2 \\
 1-{1 \over 2x} & x \ge 1/2
 \end{cases} \, ,
\end{equation}
which is obtained averaging upper and lower boundaries of $\lfloor x \rfloor / x$. The final expression for the bias is then given by:
\begin{equation}
\kappa_B(z)=\kappa_{E}(z) \; \Big(1-\chi(N_T(z))\Big)\,.
\label{eq:finalbias}
\end{equation}
We shall see in Section~\ref{sec:results} below that this simple analytic approximation can reproduce the mode of the numerically simulated magnification PDF (\ref{eq:PDF2}) for different redshifts, for different halo profiles and for different values of $N_O$. Finally, if we include the probabilistic selection effects through the parameter $\alpha$ (see Sec.~\ref{sec:selection}), we get simply
\begin{equation}
\kappa_B(z)= \kappa_{E}(z) \, \Big(1- \alpha \,\chi(N_T(z) ) \Big)\,.
\label{eq:alphabias}
\end{equation}
When $N_O$ is very large, $\chi(N_O) \rightarrow 1$ and $\kappa_B \rightarrow (1-\alpha) \kappa_E$ as expected from Eq.~(\ref{kalpha}).

\subsection{Dispersion} \label{sedispe}

In order to estimate the dispersion of the magnification PDF we first need to evaluate the streching factor Eq.~(\ref{eq:dispersion}) for our setup. We define
\begin{equation}
\omega \simeq \sqrt{\omega_r \, \omega_b}  \, ,
\end{equation}
where the factor $\omega_r$ comes from the $r_i$-summation and the factor $\omega_b$ from the $b_m$-summation. $\omega_r$ is roughly given by 
\begin{equation}
\omega_r \approx \frac{\int_0^1 dx \, g(x) \, h(x)}{\int_0^1dx \, g(x)} 
          \simeq 0.77 \,.
\end{equation}
Like the factor $E$ above, $\omega_r$ is an almost universal constant, essentially set by the form of the function $G(r,r_s)$. $\omega_b$ depends on the profile $\varphi$ and is more difficult to estimate. We find that it is reasonably well approximated by the ratio:
\begin{equation}
\omega_b \sim \frac{\Gamma (R \, Q_\varphi,t(r_s/2))}{\bar \Gamma(t(r_s/2)}\,.
\end{equation}
For example for the gaussian profile one finds $\omega_b \sim e^{-1}$ so that $\omega \simeq 0.53$.
Similarly to the evaluation of $Q_\varphi$, our derivation does not apply for singular profiles and the value of $\omega$ has to be obtained from the numerical distribution.
The essential point is again that $\omega$ can be taken a constant.

Now consider the limit $N_T\gg 1$, so that the Poisson distribution (\ref{kappaka}) can be approximated with a normal distribution. Taking into account the stretching factor $\omega$ we have (up to normalization):
\begin{equation} \label{strPDF}
P_{\rm wl}(k)  \propto \exp\left(-\frac{(k-N_{T})^{2}}{2\sigma_\omega^2} \right) \,,
\end{equation}
where $\sigma_\omega= \sqrt{N_T} \, \omega$.
The corresponding dispersion in magnitudes is found by evaluating Eq.~(\ref{kappaf}) at $k= N_T \pm \sigma_\omega$:
\begin{eqnarray}
\sigma_{\Delta m,N_O} &=& 2.5 \log_{10} 
  \frac{\sqrt{N_T(z)} - \kappa_E(z)\omega}{\sqrt{N_T(z)} + \kappa_E(z)\omega}
\nonumber \\
&\simeq& - \frac{5}{\ln 10}\frac{\kappa_{E}(z)\omega}{\sqrt{N_{T}(z)}}
\equiv{\sigma_{\Delta m, 1 \rm eff} \over \sqrt{N_{O}(z)}} \,.
\end{eqnarray}
We see that the dispersion scales as expected with $N_O$. This derivation makes sense for $\sigma_{\Delta m,N_O}$ even when $N_{C}(z) \ll 1$ given enough observations such that $N_T\gg 1$ holds. This is actually the way Ref.~\cite{Holz:2004xx} defined the effective dispersion for a singular observation, so that
\begin{equation}
\sigma_{\Delta m, 1 \rm eff} = - \frac{5}{\ln 10}\frac{\kappa_E(z) \,\omega}{\sqrt{N_{C}(z)} } \,.
\label{eq:finaldispersion}
\end{equation}
From Eq.~(\ref{eq:finalDeltaN}) it is clear that $\sigma_{\Delta m, 1 \rm eff}$ depends on the ratio $R_\varphi \equiv \omega/ Q_{\varphi}$ and not on $Q_{\varphi}$ and $\omega$ individually. Summarizing, $Q_{\varphi}$ is about the bias and $R_{\varphi}$ about the dispersion, two independent degrees of freedom.
In Table~\ref{Qphi} we list the values we will use in the next Section.
%
\begin{table}[h!]
\caption{\label{Qphi} Effective parameters $Q_\varphi $ and $R_\varphi$.}
\begin{ruledtabular}
\begin{tabular}{lccr}
Halo profile $\varphi$          & $Q_\varphi $      & $R_{\varphi}$                \\ 
\hline
Uniform  & 0.90 &  0.9  \\
Gaussian      & 0.53    & 1.3 \\
SIS   & 0.25 &  2.4 \\
\end{tabular}
\end{ruledtabular}
\end{table}

We wish to stress that Eq.~(\ref{eq:finaldispersion}) and Eq.~(\ref{eq:finalbias}) depend analytically upon the parameters of the model universe and provide an easy way to estimate magnification bias and dispersion.

\begin{figure}
\begin{center}
\includegraphics[width=8.0 cm]{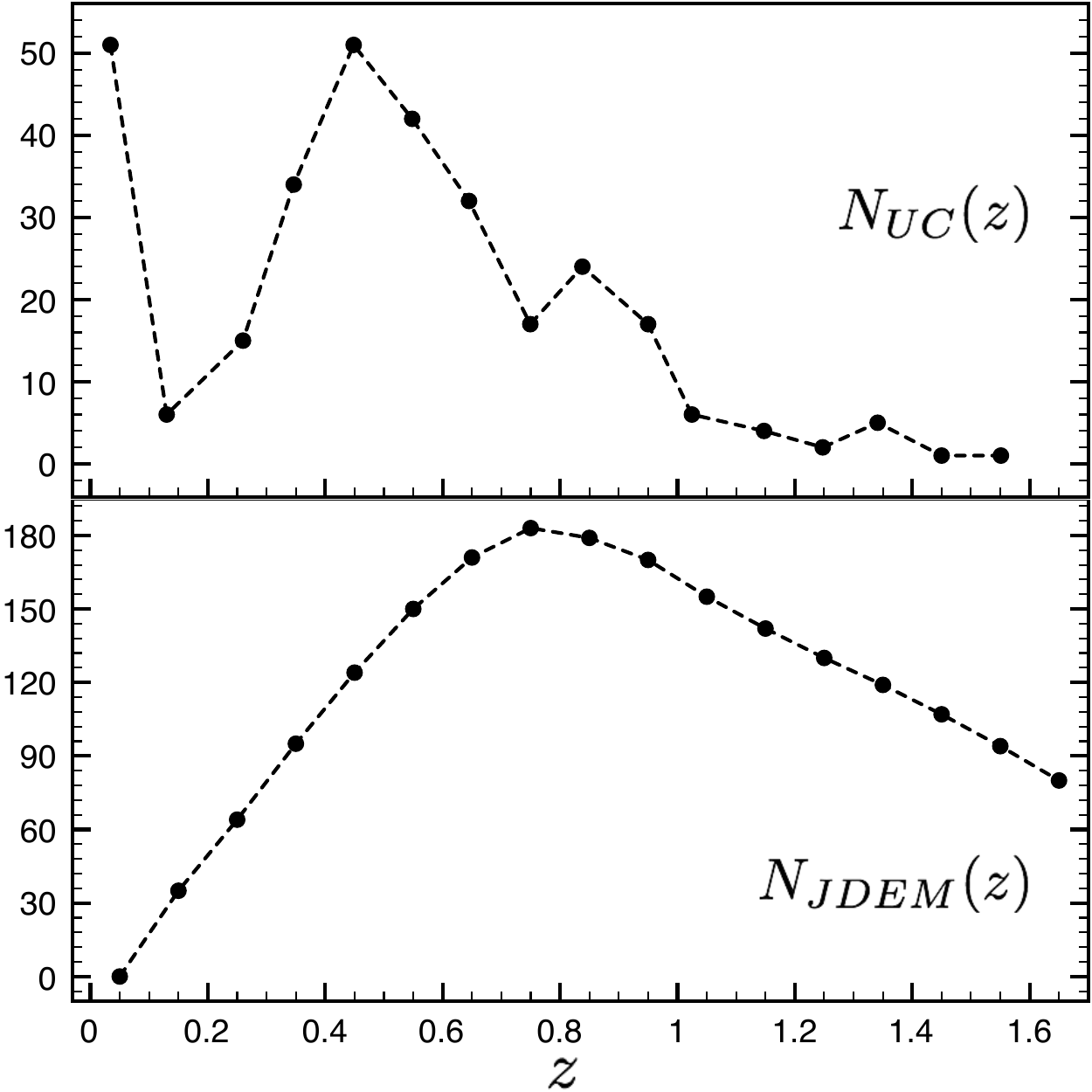}
\caption{On the top, binned number of observed SNe for $\Delta z=0.1$ for the Union Compilation of Ref.~\cite{Kowalski:2008ez} of $307$ SNe. On the bottom, binned number of observed SNe for $\Delta z=0.1$ for a JDEM-like survey of $2000$ SNe, see Ref.~\cite{Kim:2003mq}.}
\label{binned}
\end{center}
\end{figure}

Finally, note that the PDF of a single supernova is not a gaussian of standard deviation $\sigma_{\Delta m, 1 \rm eff}$ and therefore we will introduce in the next Section the full width at half maximum (FWHM) as an indicator of the dispersion for skewed distributions.
The applicability of Eq.~(\ref{eq:finaldispersion}) is indeed restricted only to data sets for which the condition $N_T(z) = N_O(z)N_{C}(z) \gg 1$ holds.
In the upper panel of Fig.~\ref{binned} we show the binned data for $N_O$ from the currently largest SNe-sample, the Union Compilation of Ref.~\cite{Kowalski:2008ez}. The data set is still quite sparse for $z \gsim 1$, which suggests using the approximation (\ref{eq:finaldispersion}) might not be applicable there to model large structures with $N_{C}(z) \lsim 1$. The lower panel in Fig.~\ref{binned} shows the simulated data for the future JDEM survey. As we shall see in Section~\ref{sec:results}, almost no bias remains in the effective PDF for the JDEM-dataset, even within a universe with very large structures.

\section{Results} 
\label{sec:results}

Here we will present numerical and analytical results for the magnification PDF.
We will first consider a model similar to the one studied by Ref.~\cite{Holz:2004xx} in order to compare our method with their ray-tracing simulations. In the next subsecion we will consider the possible effects of large-scale clustering on the magnification PDF.

Our numerical results are computed with {\tt turboGL}, a very fast Mathematica code based on the formalism developed in Sect.~\ref{wlcal}, and publicly available: on an ordinary desktop/laptop computer it takes less than a second to compute a PDF with the usual statistics of $10^{4}$ configurations. This has to be compared to expensive ray-tracing techniques.
Thanks to this performance we can use much better statistics in our analysis and simulations: $10^{5}$ configurations already give rise to very smooth histograms and curves. We would like to stress that since {\tt turboGL} is so fast, the analytic results of the previous section are not used as a tool for the analysis, but rather to provide analytic insights into the problem.

\subsection{Comparison with Holz \& Linder (2005)} \label{holzli}

Holz and Linder~\cite{Holz:2004xx} computed lensing effects in a $\Lambda$CDM universe where all matter was assumed to exist in the form of a random distribution of spherical halos the size of large galaxies.
Ref.~\cite{Holz:2004xx} used the ray-tracing simulation developed in Ref.~\cite{Holz:1997ic} from which we also obtained the value of the parameters $R_{p}$ and $d_{c}$ and the density profile used\footnote{The values of the parameters used by Ref.~\cite{Holz:2004xx} were not stated.}. $R_{p}$ is the proper radius which is assumed to be constant, while the value of $d_{c}$ fixes, through Eq.~(\ref{m1}), the halo mass. The value of the parameters used in this Section are summarized in Table \ref{tablehl}.
%
\begin{table}[h!]
\caption{\label{tablehl} Setup to be compared to the results of Ref.~\cite{Holz:2004xx}.}
\begin{ruledtabular}
\begin{tabular}{lccr}
Quantity          & Value                       \\ 
\hline
$\Omega_{M, 0}$ & $0.28$           \\
$h$             & $0.7$            \\
$d_{c}$ & $2$ Mpc           \\
$R_p$  & $200$ kpc       \\
$M_{H}$    & $1.8  \cdot 10^{12} M_{\odot}$      \\
Halo profile    & SIS      \\
\end{tabular}
\end{ruledtabular}
\end{table}
%

\begin{figure}
\begin{center}
\includegraphics[width=8.5 cm]{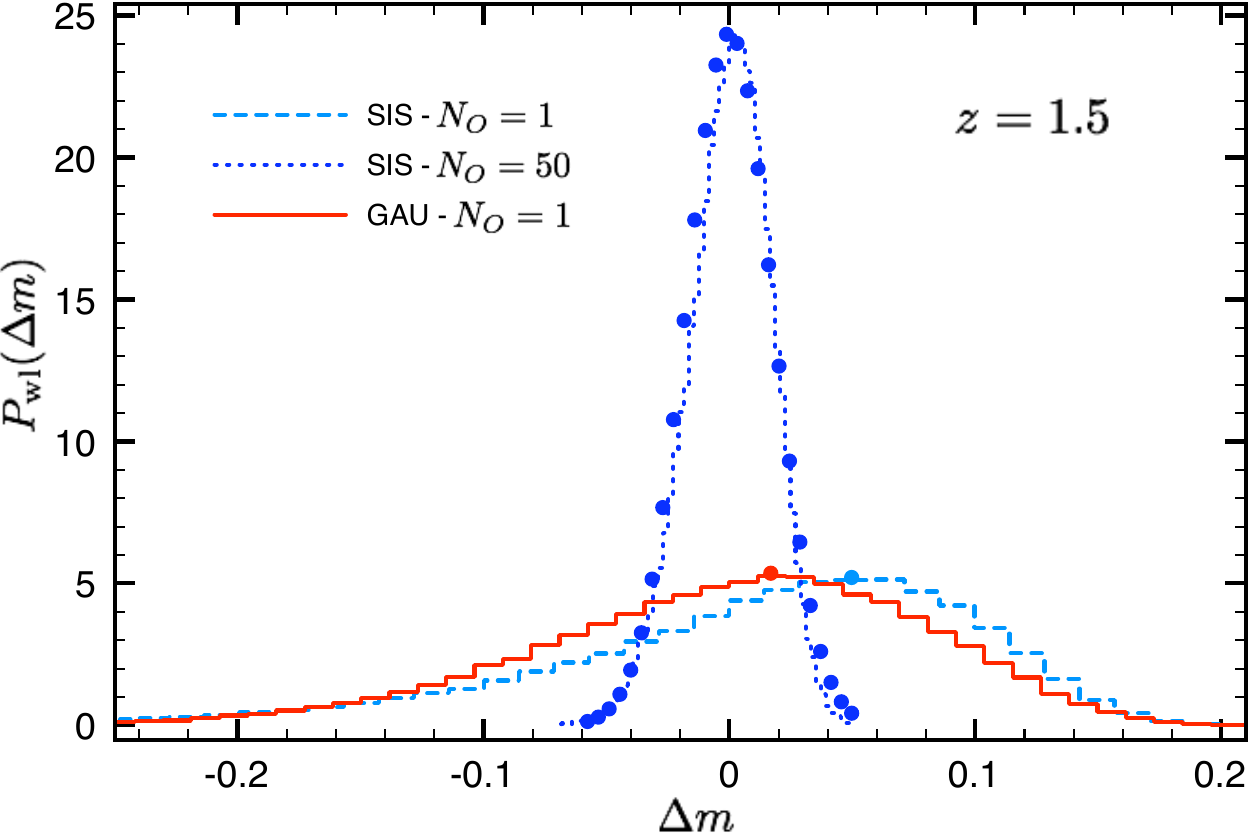}
\caption{Shown are the magnification PDFs for a source at $z=1.5$ for the $\Lambda$CDM model of Table \ref{tablehl} with SIS and gaussian halo profiles. The SIS case has been evaluated for both $N_{O}=1$ and $N_{O}=50$ SNe measurements. Each histogram has a statistics of $10^{5}$ realizations.
Also shown (filled circles) are the approximate PDFs given by the distribution of Eq.~(\ref{strPDF}). For the PDFs relative to $N_{O}=1$ only the mode of the distribution is shown.}
\label{PDFs}
\end{center}
\end{figure}

In Fig.~\ref{PDFs} we plot the magnification PDF for a source ($N_{O}=1$) at redshift $z=1.5$ in the case of SIS and gaussian halo profiles. Both PDFs are visibly skewed, but the bias is still realatively weak, thanks to the high degree of homogeneity given by the parameters of Table \ref{tablehl}. The SIS-PDF has both a larger bias and a more pronounced tail at high-magnifications than does the gaussian PDF. This was expected because the strong cusp of the SIS leads to more high-magnification events compensated by a drop at small magnifications.
Also shown in Fig.~\ref{PDFs} is the SIS-PDF relative to a set of $N_{O}=50$ SNe measurements: as expected skewness and dispersion are reduced showing that, as discussed in Section \ref{sec:observations}, the distribution approaches a gaussian for large $N_{O}$, eventually reducing to a $\delta$-function at zero convergence. These results are in good agreement with Fig.~1 of  Ref.~\cite{Holz:2004xx}, which shows that our statistical weak lensing approximation can well reproduce the results of ray-tracing calculations. In Fig.~\ref{PDFs} we also show (filled circles) the analytic effective probability distribution developed in Section~\ref{sec:analytic}.
As was discussed in subsection~\ref{sedispe}, the effective PDF distribution is a useful quantity only for $N_{T} \gg 1$, and so we plotted the full effective PDF only for the SIS case with $N_{O}=50$, where $N_{T}\simeq 37$. The agreement with the exact results is remarkably good. For the $N_{O}=1$ cases the point density is not sufficient to model the full PDF, but the modes of the distributions are still well produced by the analytic approximation (single filled circles on $N_O=1$ curves).

Besides the standard deviation and similarly to Ref.~\cite{Holz:2004xx}, we will use the full width at half maximum (FWHM) as an alternative indicator for the dispersion of a skewed distribution.
In particular we define three different indicators:
\begin{eqnarray} \label{fwhmd}
v_{-}&=& {\textrm{FWHM}_{-} \over 1.18} \nonumber \\
v_{+}&=& {\textrm{FWHM}_{+} \over 1.18} \nonumber \\
v&=& {v_{-}+v_{+} \over 2} = {\textrm{FWHM} \over 2.35} \, ,
\end{eqnarray}
where FWHM$_{-/+}$ are defined as the distances from the left/right edges to the mode of the total FWHM and so $\textrm{FWHM}_{-}+ \textrm{FWHM}_{+}= \textrm{FWHM}$ (see Fig.~\ref{PDFs2} for an illustration).
We found these indicators particularly useful because for a gaussian distribution one has $v_{-}=v_{+}=v=\sigma$ where $\sigma$ is the standard deviation and so departures from this limit signal the presence of skewness in the distribution.
Mode together with $v_{-/+}$ will describe the skewed peak of a PDF, while $v$ and $\sigma$ its dispersion.
The idea is that for skewed distributions and small datasets $v$ is more meaningful than $\sigma$, which is affected by the long high-magnification tail which is of little importance if we deal with small data sets.
This is actually the reason why we focused on the FWHM to characterize the skewness: the third standardized moment (the skewness $\gamma_{1}$), for example, would have been again sensitive to the long and low high-magnification tail.

\begin{figure}
\begin{center}
\includegraphics[width=8.5 cm]{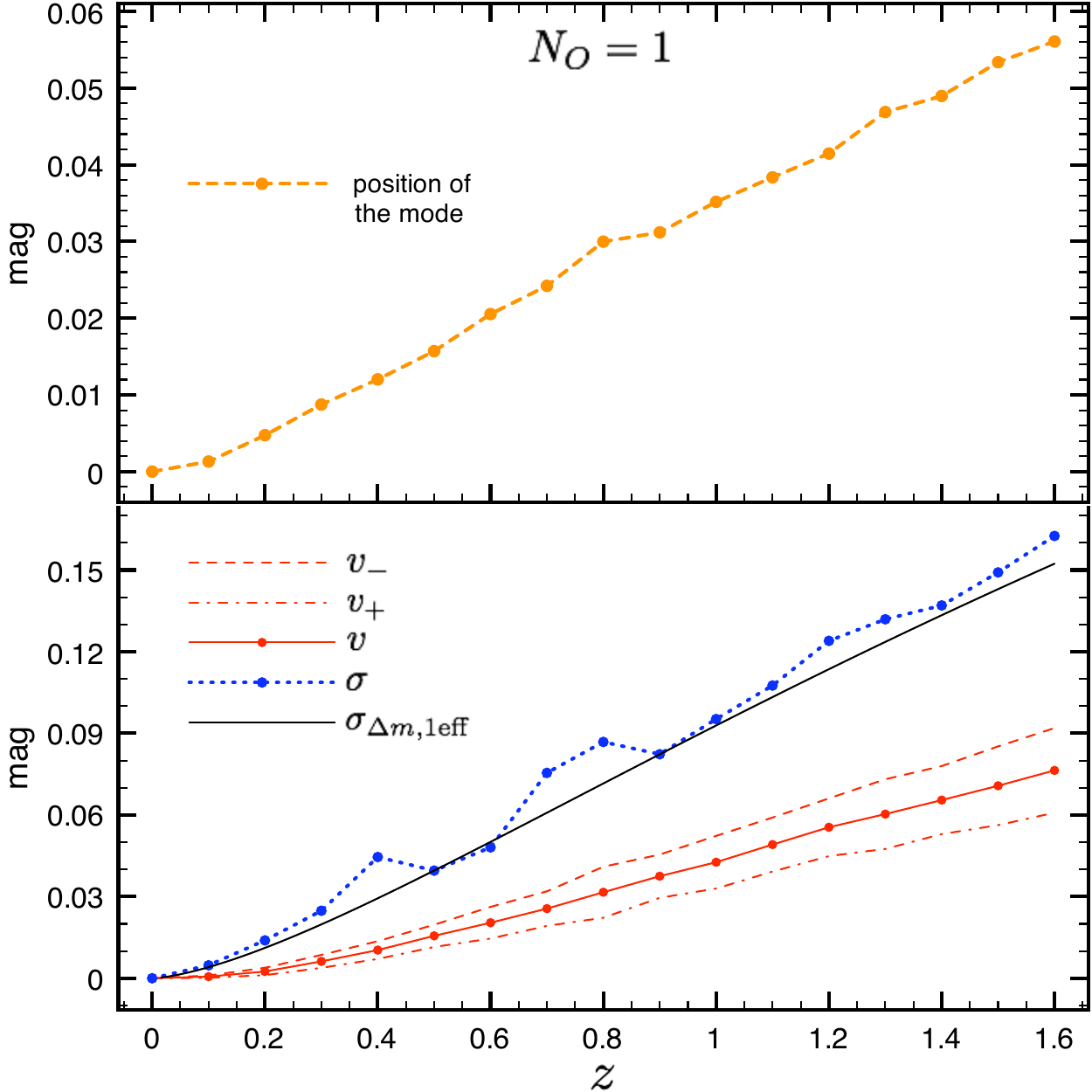}
\caption{Shown are the bias (top) and the dispersion (bottom) of the magnification PDF in the $\Lambda$CDM model of Table \ref{tablehl} as a function of the redshift of the source for $N_O=1$. Filled circles mark the values obtained from the numerical simulation (omitted for clarity in the $v_{-}$ and $v_{+}$ curves). $\sigma_{\Delta m, 1 \rm eff}$ is calculated from Eq.~(\ref{eq:finaldispersion}).}
\label{modi1}
\end{center}
\end{figure}

In Fig.~\ref{modi1} we show the bias and the dispersion for the SIS profile for $N_{O}=1$. This figure can be compared with Figs.~5 and 7 of  Ref.~\cite{Holz:2004xx}. While the bias introduced by the skewness of the distribution is moderate as shown from the top panel of Fig.~\ref{modi1}, there is a clear deviation from gaussianity, as one finds that $\sigma \sim 2 \, v$. Moreover $\sigma_{\Delta m, 1 \rm eff}$ very well reproduces the numerical results for $\sigma$ and consequently overestimates the dispersion $v$.

\begin{figure}
\begin{center}
\includegraphics[width=8.5 cm]{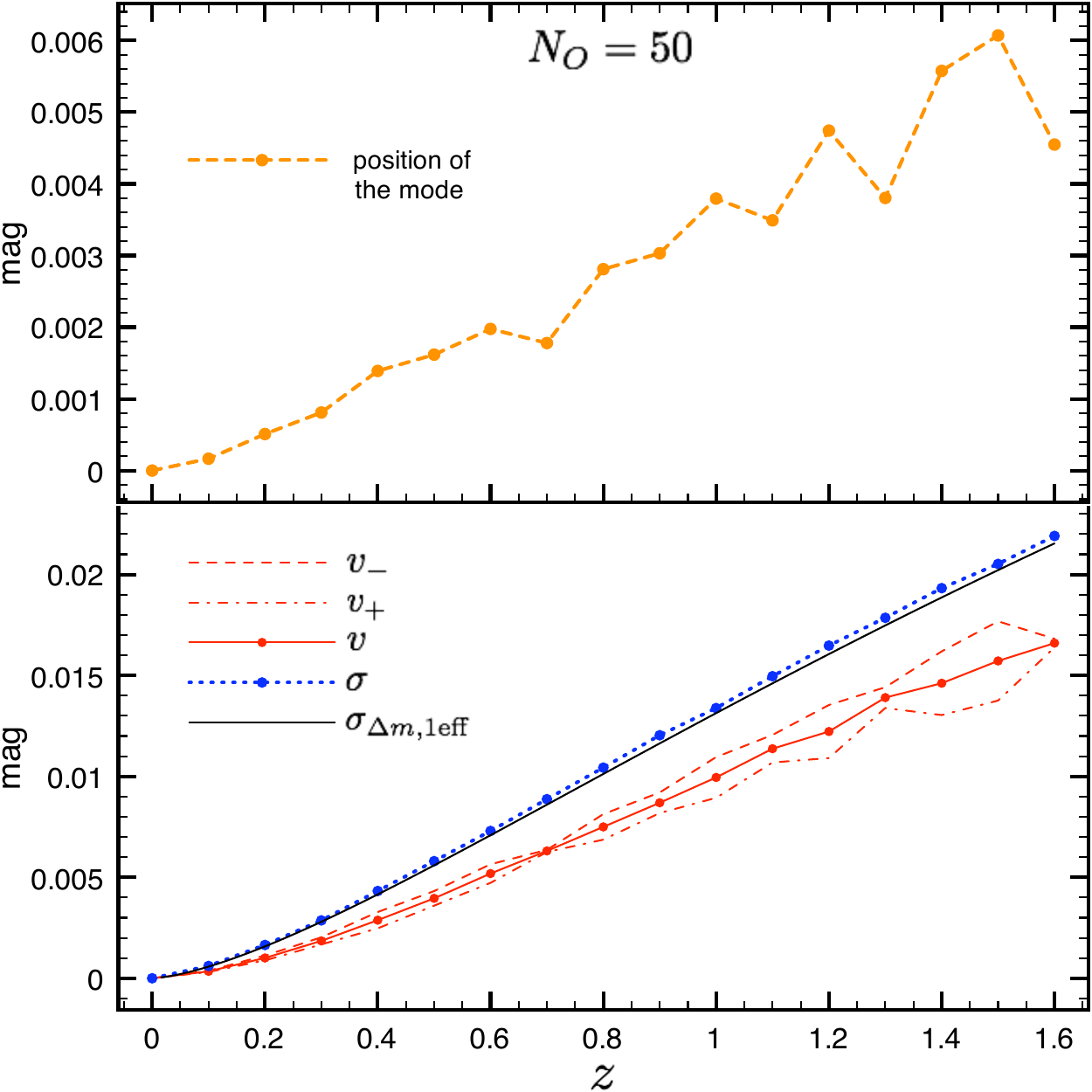}
\caption{Same as Fig.~\ref{modi1}, but for $N_O=50$. The dispersions $v$ and $\sigma$ are now closer than in Fig.~\ref{modi1}  showing that, by increasing the number of observations, the lensing PDF approaches a gaussian.}
\label{modi2}
\end{center}
\end{figure}

Fig.~\ref{modi2} is similar to Fig.~\ref{modi1} but now constructed for $N_{O}=50$. The bias introduced by the skewness of the distribution is now greatly reduced. Moreover the dispersions $v$ and $\sigma$ are converging ($\sigma \sim 1.3 \, v$) showing that averaging over a  large sample of measurements can reduce the effects of the skewness of the distribution.

\begin{figure}
\begin{center}
\includegraphics[width=8.5 cm]{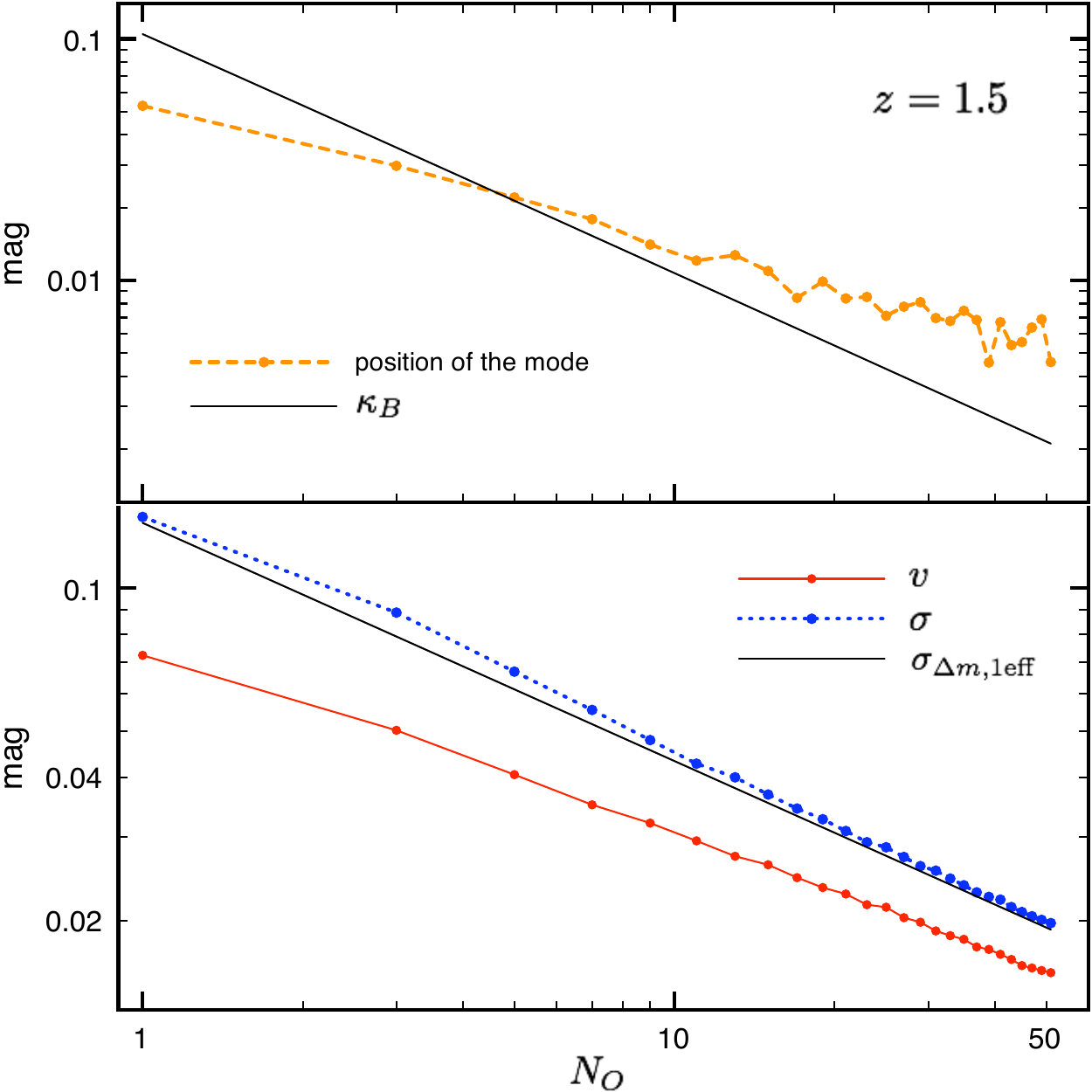}
\caption{Shown are the bias (top) and the dispersion (bottom) of the magnification PDF in the $\Lambda$CDM model of Table \ref{tablehl} as a function of the size $N_O$ of the data set for a source at redshift $z=1.5$. Filled circles mark the values obtained from the numerical simulation. $\sigma_{\Delta m, 1 \rm eff}$ is calculated from Eq.~(\ref{eq:finaldispersion}) and $\kappa_{B}$ from Eq.~(\ref{eq:finalbias}).}
\label{modi3}
\end{center}
\end{figure}

In order to investigate the convergence of the lensing PDF to a gaussian, we studied in Fig.~\ref{modi3} the dependence of bias and dispersion upon the data set size $N_{O}$.
The bias (top panel of Fig.~\ref{modi3}) is plotted together with the analytic result of Eq.~(\ref{eq:finalbias}):
$\kappa_{B}$ scales like $1/N_{O}$ and so we conclude that the actual bias converges slower than that, in agreement with Fig.~2 of Ref.~\cite{Holz:2004xx}.
The bottom panel of Fig.~\ref{modi3} instead shows that $\sigma$ is well described by $\sigma_{\Delta m, 1 \rm eff}$, which means that $\sigma$ scales, as expected, like $1/\sqrt{N_{O}}$.
This plot agrees again with Fig.~4 of Ref.~\cite{Holz:2004xx} and shows that $\sigma$ converges to $v$ only for very large $N_{O}$.
We stress at this point that this strong non-guassian signature is not a general feature but is due to the strong cusp of the SIS profile and is largely reduced with other density profiles like the NFW or the gaussian.

As a technical note, we point out that to obtain the non-gaussian signature of Fig.~\ref{modi3}, we had to increase the binning in the impact parameter in order to accurately resolve the cusp.
This of course increases computational time and so it is worth asking what is the impact of a less accurate binning.
Our conclusions are that losing that non-gaussian signature is not an important problem because the skewness of the distribution is {\em under}estimated only when the latter is already very close to a gaussian and thus at worst leads to a slight conservative {\em underestimation} of the bias for large data sets and SIS profiles.

The results of this Section  show that our statistical model, based on cumulative weak lensing, very well describes the most important effects of lensing by the inhomogenous matter distributions. However, the model discussed in this Section, defined by the parameters in Table~\ref{tablehl},  neglects all effects of large scale clustering on the weak lensing. Such clustering is nevertheless clearly present in the form of large voids, superclusters and filamentary structures seen both in the large scale simulations and in the actual redsift survey datasets. We will next try to get an idea of the potential impact of these effects by considering a universe made out of much larger halos.

\subsection{Large-scale structures}

In this Section we study the lensing effects caused by very large-scale structures. For simplicity we still considered universe models made of a statistical distribution of a single type of halos. We performed several simulations with different halo masses and found that the bias effects start to become significant when the halo masses are at least of the order of the largest gravitationally bound superclusters: $M \gsim  10^{14} h^{-1} M_{\odot}$.  However, even larger nonvirialized structures can exist and to illustrate their effects we chose models with even larger masses. The details of the models are given in Table~\ref{table}.
%
\begin{table}[h!]
\caption{\label{table} Parameters modelling large-scale structures.}
\begin{ruledtabular}
\begin{tabular}{lccr}
Quantity          &$\Lambda$CDM    & EdS                        \\ 
\hline
$\Omega_{M, 0}$ & $0.28$ & $1$              \\
$h$             & $0.7$ & $0.5$   \\
$d_{c}$ & $15 \, h^{-1}$ Mpc         & $15 \, h^{-1}$ Mpc   \\
$R_p$  & $1.1 \, h^{-1}$ Mpc    & $1.1 \, h^{-1}$ Mpc   \\
$M_{H}$    & $1.5  \cdot 10^{15} h^{-1} M_{\odot}$    & $5.5 \cdot 10^{15} h^{-1} M_{\odot}$  \\
Halo profile    & gaussian & gaussian     \\
\end{tabular}
\end{ruledtabular}
\end{table}
We will consider both the $\Lambda$CDM model with $\Omega_{M, 0}=0.28$ and $h=0.7$ and the EdS model with $\Omega_{M, 0}=1$ and $h=0.5$. The latter has been given a lower Hubble constant in order to better agree with the CMB observations. For simplicity, we are again taking our objects to have a constant proper radius $R_p$, although for large, possibly unvirialized structures this is not necessarily true. We shall come back to this issue in a forthcoming publication \cite{km3}. The size of the halos in Table~\ref{table} roughly corresponds to the scale of large superclusters and has been chosen in order to have an average density contrast of 200 at virialization which we assume to happen at $z=1.6$.
For cluster scale halos the most appropriate profile to be used would be the NFW profile~\cite{Navarro:1995iw}.
In this study, however, we have used for simplicity the gaussian profile which does not need any extra parameter. Again we will investigate various profiles in a forthcoming publication \cite{km3}.

\begin{figure}
\begin{center}
\includegraphics[width=8.5 cm]{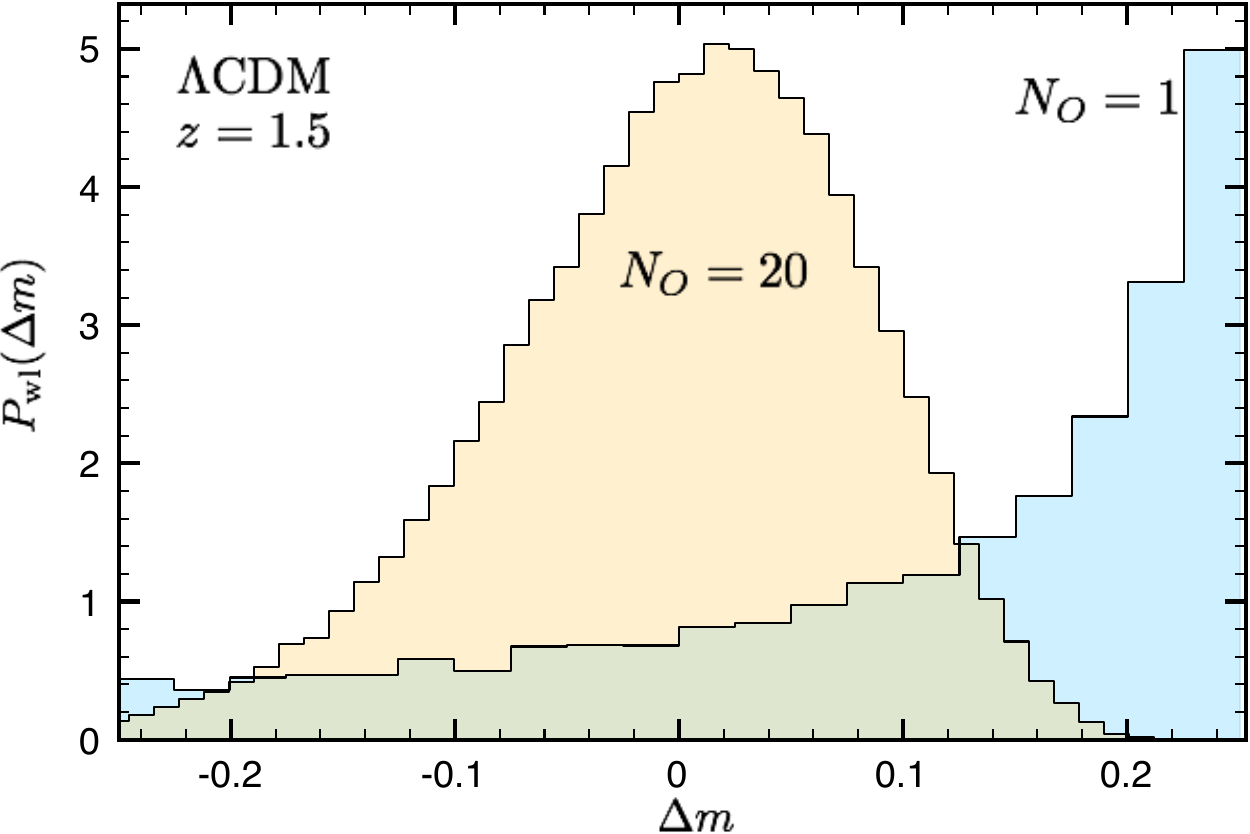}
\caption{Shown are the magnification PDFs for the $\Lambda$CDM model described in Table \ref{table} for a source at $z=1.5$ in the case of $N_{O}=1$ and $N_{O}=20$ observations.}
\label{PDFs3}
\end{center}
\end{figure}

\begin{figure}
\begin{center}
\includegraphics[width=8.5 cm]{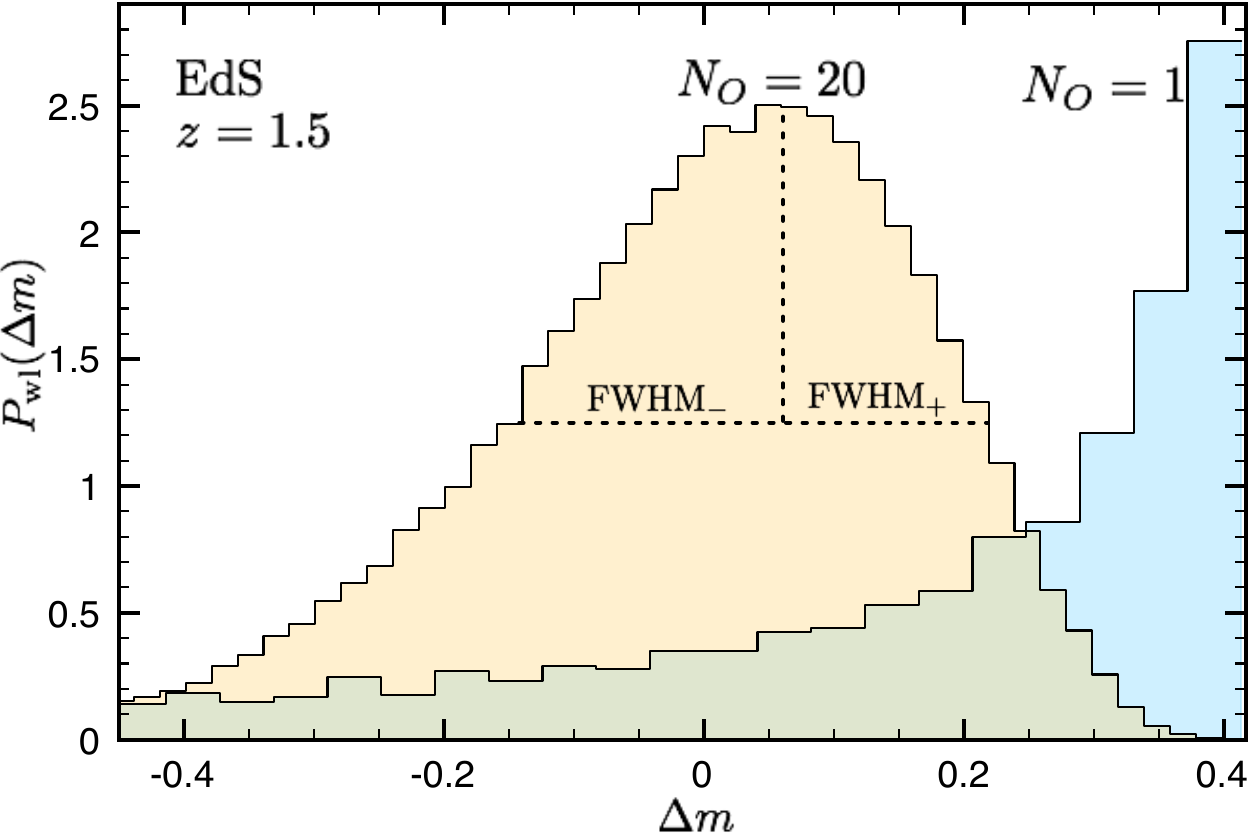}
\caption{Same as in Fig.~\ref{PDFs3}, but now for the EdS-model described in Table~\ref{table}.
Also displayed is the definition of FWHM$_{-}$ and FWHM$_{+}$ used in Eqs.~(\ref{fwhmd}).}
\label{PDFs2}
\end{center}
\end{figure}

\begin{figure*}
\begin{center}
\includegraphics[width=14 cm]{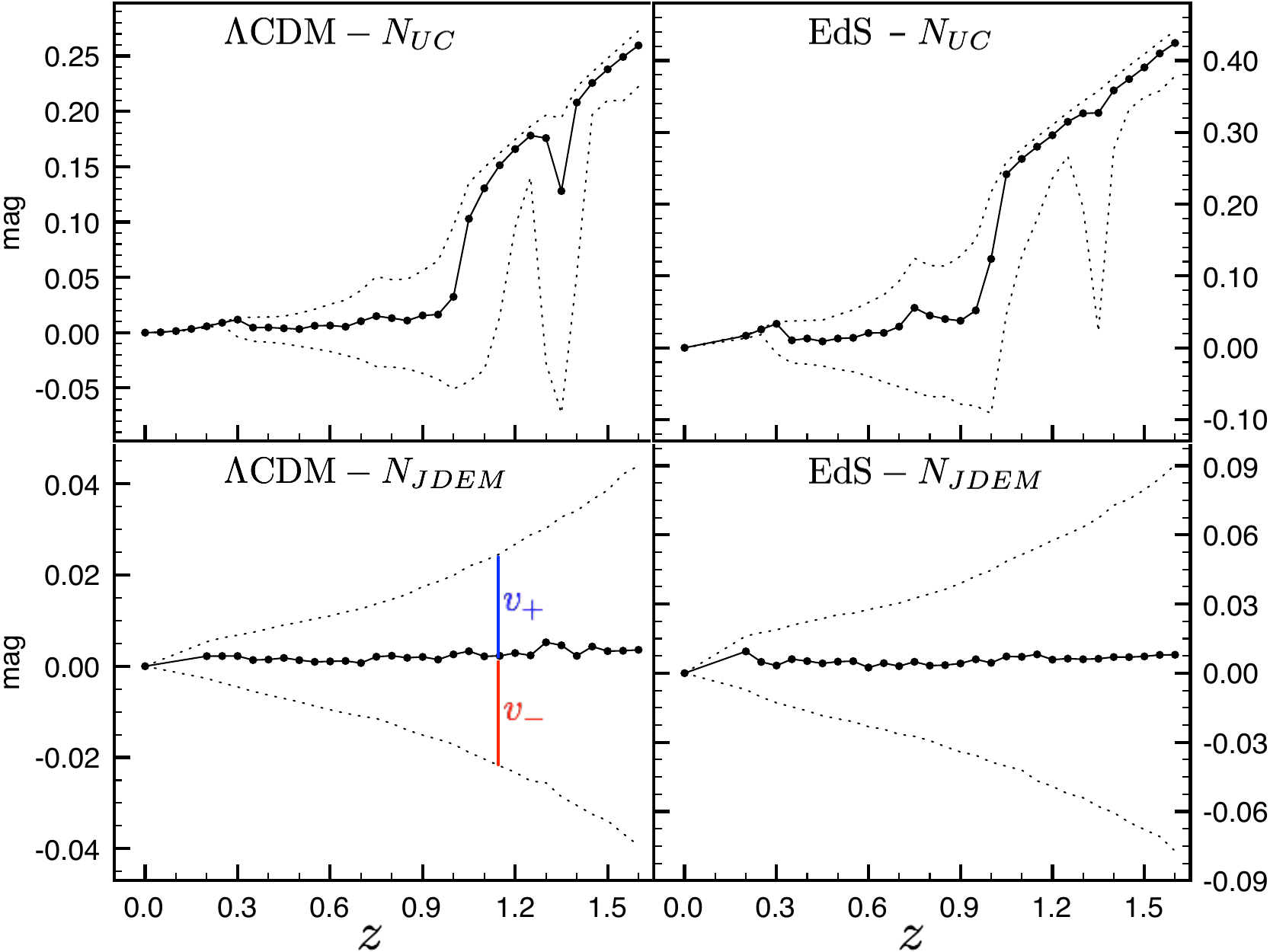}
\caption{Lensing bias as a function of redshift for the parameters of Table \ref{table}. The left panels are for the $\Lambda$CDM model, while the right panels are for the EdS model. Moreover the top panels use the redshift distribution for SNe measurements given by the Union Compilation, while the bottom panels use the redshift distribution for a JDEM-like survey of $2000$ SNe, see Fig.~\ref{binned}.
Filled circles mark the values obtained from the numerical simulation and the dotted curves give the dispersions $v_{-}$ and $v_{+}$, as indicated in the figure.}
\label{LS44}
\end{center}
\end{figure*}

We plot the magnification PDF for the models of Table~\ref{table} in Figs.~\ref{PDFs3}-\ref{PDFs2} for the $\Lambda$CDM and for the EdS models, respectively. As expected, the distributions are now significantly more skewed than they were in the case of the smaller halos of the previous Section: in both cases the fundamental PDF ($N_{O}=1$) has the mode at the demagnification corresponding to the empty beam, the maximum demagnification possible.
Moreover Figs.~\ref{PDFs3}-\ref{PDFs2} clearly illustrate the importance of introducing the dispersion indicators of Eqs.~(\ref{fwhmd}). Even if the $\sigma$ relative to the PDF of $N_{O}=1$ is larger than the $\sigma$ relative to the PDF of $N_{O}=20$ because of the long high-magnification tail, the dispersion $v$ behaves in the opposite way showing that with little data sets the actual dispersion is much smaller than the one computed assuming almost-gaussian distributions.

In Fig.~\ref{LS44} we show the magnification bias $\Delta m$. Left panels correspond to the $\Lambda$CDM and right panels to the EdS model as the GBS. As we have explained, the bias of the effective PDF depends quantitatively on the number of observations at each redshift. The upper panels show the bias computed from $N_O$ corresponding to the UC-data displayed in the upper panel of Fig.~\ref{binned}. The line has been made continuous by using a continuous intepolation for $N_O(z)$ between the actual data points. The lower panels show the results for $N_O$ corresponding to the the simulated data of the JDEM experiment. Also plotted are the dispersions $v_{-}$ and $v_{+}$ introduced in Eqs.~(\ref{fwhmd}): they give a quantitative estimation of the dispersion relative to the plotted bias. We stress that $v_{-}$ and $v_{+}$ refer to the PDF relative to $N_O$ measurements and therefore they are good indicators of the actual dispersion relative to the data set used.

First note that the EdS model has larger lensing effects (observe that in Fig.~\ref{LS44} the scales in the left and the right panels are not the same). This is clearly seen from the convergence for the empty beam given in Eq.~(\ref{eq:kappa3}): $\kappa_{E} \propto h^{2} \, \Omega_{M, 0}$ and even if the EdS universe has a lower $h$, the larger value of $\Omega_{M, 0}$ is dominant: $\left. h^{2} \, \Omega_{M, 0} \right|_{\textrm{EdS}}  /  \left. h^{2} \, \Omega_{M, 0} \right|_{\Lambda\textrm{CDM}} \sim 1.7$.
See Ref.~\cite{Kainulainen:2009sx} for an application of the latter amplification of lensing effects in the EdS model.

Moreover, the Union Compilation of Ref.~\cite{Kowalski:2008ez} does not have enough SNe measurements at high redshifts ($z>1$) to suppress the lensing bias, while a JDEM-like survey of $2000$ SNe will be able to recover the GBS result within the intrinsic SNe dispersion of $\sigma_{M}\sim 0.1$ mag, that is, $\Delta m \ll \sigma_{M}$. If the model universe of Table \ref{table} captures the actual degree of inhomogeneity of the universe, then these results suggest that lensing bias has to be incorporated within the data analysis of SNe observations for the present-day data sets.
Let us remind that the crucial feature that our model universe captures, and which gives rise to the large biases, is that photons can travel through voids and miss the localised overdensities.  This feature is absent from swiss-cheese models where the boundaries of the holes are designed to have compensating overdensities. In these models a photon that passes through a void always has to pass also through a compensating high-density shell, which results in a constrained PDF. It is not surprising then that such models have been shown to have on average little lensing effects \cite{Brouzakis:2007zi,Vanderveld:2008vi}.

\section{Conclusions} \label{conclusions}

In this paper we have presented a new method to calculate the magnification probability distribution function (PDF) for a universe made of randomly distributed halos. The method is based on the weak lensing approximation and on generating stochastic configurations of halos along the line of sight, or along the photon geodesic from source to the observer. The basic physical feature incoporated by the method is the fact that underdensities occupy most of the volume while most of the mass lies in overdense regions in the form of clusters and filaments. As a consequence the column density along a single geodesic is likely to be lower than the average density of the global background solution. This is of potential importance for the interpretation of the supernova observations, which are still probing much smaller angular scales than the scale at which the homogeneity of the GBS is recovered.

We derived a simple statistical formula by use of which one can easily compute the fundamental PDF for a single event and the effective PDF for quantities averaged over a number of observations $N_O$. Our formulae were explicitly written for arbitrary mass distributions of spherical halos with arbitrary density profiles and they can be straightforwardly extended to include also different halo geometries (say thin cylinders to model filaments). We also showed how one can easily incorporate  most selection effects into the formulae for the observable fundamental PDF. Along with this paper, we  released the {\tt turboGL} package~\cite{turbogl}, a simple and very fast Mathematica code to perform numerical simulations based on our model. The code will be continuously updated to incorporate more features (halo mass distributions, geometries, systematic biases etc) in the future.

While our method can easily compute the PDF relative to arbitrary halo mass distributions, selection biases and halo profiles and evolutions, we focused for simplicity on the simplest configuration of one family of halos with no evolution and no selection biases in the numerical examples shown in the present paper. More general results will be presented in a forthcoming paper \cite{km3}. We checked the validity of our weak lensing approximation against exact results of light propagation in an inhomogeneous universe and the validity of our stochastic approach against the ray-tracing simulations of Ref.~\cite{Holz:2004xx} (in a model universe with halos of mass $M\sim 10^{12} h^{-1} M_\odot$). In both cases we were able to recover the main results within a few percent accuracy.

In addition, we considered the biasing effect on the PDF due to very large structres, $M\sim 10^{15} h^{-1} M_\odot$, whose existence is suggested by the large voids and filamentary structures seen both in the large-scale simulations and in the galaxy redshift surveys~\cite{BoylanKolchin:2009nc,Fosalba:2007mf,surveys}. We produced simulated PDFs for such universes both in the $\Lambda$CDM and in the Einstein-de Sitter (EdS) background model. We also produced the distributions for binned sets of observations and computed the biases and dispersions for these effective PDF. Our results suggest that the lensing bias could affect the extraction of cosmological parameters from the current best data sets. However, we found that a JDEM-like survey should be able to remove the lensing effects of even the largest imaginable structures assuming, of course, that no {\em selection} biases were present in the measured SNe.  

Indeed, in addition to computing the statistical bias, and perhaps beyond that, the most potential use of our method may be in computing the effect of different sets of selection biases on the  observational magnification PDF. For example, the effects of rejecting outliers, or existence of zones of avoidance in the sky, possibly correlated with the densest structures, are easily incorporated. This issue will be matter of further investigation \cite{km3}. The method can be extended to include also strong lensing effects and eventually it will be interesting to incorporate also the redshift effects to our simulation package.

\appendix

\section{Partially filled beams in EdS universe} 
\label{weds}

Here we shall compare the analytic expressions for exact and cumulative weak lensing convergence functions for empty and partially filled beams in the EdS model. Given the redhsift dependence of the co-moving distance
\begin{equation}
r(z) = \frac{2 c}{a_0 H_0} \left(1-(1+z)^{-1/2} \right) 
\label{rz} 
\end{equation}
we can compute the empty bin demagnification of a source at redshift $z$ 
($r_s = r(z)$)
\begin{eqnarray}
\kappa_{E}(z)&=&-{3 \over 2}{a_{0}^{2}H_{0}^{2}\over c^{2}}
\int_{0}^{r_{s}}dr{r(r_{s}-r) \over r_{s}}{a_{0} \over a(t(r))}
\nonumber \\
&=& 12 -3 \,{1+(1+z)^{-1/2} \over 1-(1+z)^{-1/2}} \ln (1+z) \,.
\end{eqnarray}
Note that $\kappa_{E}(z)$ does not depend on $H_{0}$. That is, it only depends on the redshift difference and not on the time of the observation.
For a partially filled beam, with the filling factor
\begin{equation}
  \alpha= {\rho_{\rm beam} \over \bar{\rho}},
\label{App:alpha}
\end{equation}
the weak lensing result becomes simply
\begin{equation} \label{lino}
\kappa^{wl}_{E, \alpha}(z) =(1-\alpha) \, \kappa_E(z) \,,
\end{equation}
as shown in Section \ref{sec:selection}.
Let us now compare this result to the exact formulae from Refs.~\cite{zel,dashe66,Kantowski:1969}. Zel'Dovich~\cite{zel} was perhaps the first to recognize the importance of inhomogeneities on FLRW distances. He found the following exact result for the luminosity distance along an empty beam (a line of sight emptied of any matter) in the EdS universe:
\begin{equation} 
D_{L, \, 0}(z) = \frac{2c}{5H_0} \, (1+z)^{2} \left(1-(1+z)^{-5/2} \right) \,.
\label{re1}
\end{equation}
This result can be extended to the case of a partially filled beam~\cite{dashe66}:
\begin{equation} 
D_{L, \, \alpha}(z) = \frac{2c}{kH_0} \, (1+z)^\frac{k+3}{4} \left(1-(1+z)^{-k/2} \right),
\label{re2}
\end{equation}
where:
\begin{equation} \label{alpha2}
k=(25-24 \, \alpha )^{1/2}\,,
\end{equation}
with $\alpha$ given in Eq.~(\ref{App:alpha}). Note that for $\alpha=0$ Eq.~(\ref{re2}) reduces to Eq.~(\ref{re1}) and for $\alpha=1$ it reduces to the EdS luminosity distance: $D_{L,1}(z) = (1+z)a_{0}r(z)$, where $r(z)$ is given by Eq.~(\ref{rz}). Using Eq.~(\ref{rz}) and Eq.~(\ref{re2}) we can then compute the (de)magnification $\mu = (D_{L,\alpha}/D_{L,1})^2$ and the convergence:
\begin{eqnarray}
\kappa^{ex}_{E,\alpha}(z)&=&1-\mu_{\alpha}^{-1/2}
\nonumber \\
&=& 1-{(1+z)^{(k-1)/4} \over k} {1- (1+z)^{-k/2} \over 1- (1+z)^{-1/2}} \,. \;
\label{app:exact}
\end{eqnarray}
Results (\ref{lino}) and (\ref{app:exact}) look very different, but they agree numerically to within a few per cent up to $z\approx 2$, as can be seen from the Fig.~\ref{check}. For small $z$ the agreement of (\ref{lino}) and (\ref{app:exact}) can be seen explicitly from their power series expansions:
\begin{equation}
\kappa^{wl}_{E,\alpha}(z) = \frac{1-\alpha}{4} 
  \Big(- z^2 + z^3 - \frac{73}{80} z^4 \Big) + {\cal O}(z^5) 
\end{equation}
and
\begin{equation}
\kappa^{ex}_{\alpha}(z) - \kappa^{wl}_{E,\alpha}(z) = -\frac{3(1-\alpha)^2}{160}z^4 
+ {\cal O}(z^5) \,.
\end{equation}
The first two terms of the expansions vanish in accordance with the fact that lensing is negligible at small redshifts. Moreover, the functions  $\kappa^{ex}_{\alpha}$ and $\kappa^{wl}_{E,\alpha}$ agree exactly up to third order and up to the fifth order the correction is $\sim (1-\alpha)^2$. That is, while for $\alpha \rightarrow 1$ the convergence goes to zero as $1-\alpha$, the difference between the exact result and the weak lensing approximation goes to zero even faster $(1-\alpha)^{2}$.

\section{Identity of Eq.~(\ref{eq:identity})} 
\label{identity}

We want to express the following integral $I$ as a volume integral of $\varphi$:
\begin{equation}
I=2 \pi \int_{0}^{R} db \, b \, \Gamma(b, t)
\equiv 2 \pi \int_{0}^{R} db \,  b\int_{b}^{R} {2 \, l \, dl \over \sqrt{l^{2}-b^{2}}} \varphi(l,t) \nonumber .
\end{equation}
First we rewrite the latter as
\begin{equation} \label{integ2}
I=2 \pi \int_{0}^{R} db \,  \int_{b}^{R} {d \over db} \left(-\sqrt{l^{2}-b^{2}} \right) \, 2\, l \, dl \, \varphi(l,t) \, .
\end{equation}
Then we introduce the auxiliary function
\begin{equation}
Z(b)=-\int_{b}^{R} \sqrt{l^{2}-b^{2}} \, 2\, l \, dl \, \varphi(l,t) \nonumber \, ,
\end{equation}
and calculate its derivative:
\begin{eqnarray}
Z'(b)&=& \left. \sqrt{l^{2}-b^{2}} \, 2\, l \, dl \, \varphi(l,t) \right|_{l=b} \nonumber \\
&- & \int_{b}^{R} {d \over db}\sqrt{l^{2}-b^{2}} \, 2\, l \, dl \, \varphi(l,t) \nonumber \\
&=&0+\int_{b}^{R} {d \over db} \left(-\sqrt{l^{2}-b^{2}} \right) \, 2\, l \, dl \, \varphi(l,t)  \nonumber \, .
\end{eqnarray}
Finally we substitute the latter expression in Eq.~(\ref{integ2}):
\begin{eqnarray}
I&=&2 \pi \int_{0}^{R}  db \,  Z'(b)  =2 \pi (Z(R)-Z(0)) \nonumber \\
&=& 4\pi \int_{0}^{R(t)} db \, b^{2} \, \varphi(b,t)  \nonumber \, .
\end{eqnarray}
%



\end{document}